# Analysis of Hematocrit-Plasma Separation in a Trifurcated Microchannel by a Diffusive Flux Model

Rishi Kumar[1], Indranil Saha Dalal[1*], and K. Muralidhar[2*]

[1]Department of Chemical Engineering, Indian Institute of Technology Kanpur, India 208016

[2]Department of Mechanical Engineering, Indian Institute of Technology Kanpur, India 208016

**Abstract**

Platelet-enriched plasma and red blood cells (RBC) are needed in the treatment of blood-related diseases, including anaemia and blood cancer. These essential components must be separated from blood in well-designed experimental setups. If active techniques are used, the blood components are likely to be damaged or contaminated while handling. Passive techniques for component separation are preferred, and their design for effectiveness before manufacturing is the subject of this article. Specifically, the performance of a design consisting of a trifurcated microchannel is examined in the framework of 3D numerical simulation, following similar design ideas in recent experimental studies. The influence of geometrical parameters of the channel, such as width and separation arm angle, inlet extension, flow constriction, and flow parameters, including flow rates, hematocrit concentration, and temperature, is studied. The present study utilizes the diffusive flux model (DFM) to model the shear-driven migration of red blood cells (RBC) in a microchannel along with an appropriate rheology model. The physical mechanism driving separation is the formation of the cell-free layer near the walls, using which the separation efficiency and device effectiveness are quantified. It is found that a microchannel with a smaller width and an extended inlet, along with diluted blood samples of lower hematocrit, is effective for greater separation, while the device performance is less sensitive to the flow rates, flow constriction, and the separator angle.

## 1. Introduction

Blood is a fluid mixture of various constituents essential to support life. The background liquid is plasma containing suspended particulates, including white blood cells (leukocytes), platelets, proteins, and RBCs. The former, i.e., RBCs, are soft biconcave disks[1] and form the most abundant constituent of blood, accounting for 40-45% of its volume. Treating diseases such as blood cancer and anaemia requires platelet-enriched plasma, depleted in RBCs. To assist in treatment, these are individually added to the blood and thus, as a first step, require that they be separated from healthy blood. Separation techniques such as centrifugation, dilution, fractionation, and dialysis play a crucial role in this regard. In a separation process, RBCs are extracted, leaving behind a suspension rich in platelets. In centrifugation, denser and heavier particles migrate to the edge. This process is time-consuming and complex, and high centrifugal forces can damage the constituents through a process such as haemolysis. Other active devices utilize electrical, magnetic, acoustic, optical, and gravitational forces. The development of passive devices that are based on microfluidic techniques is a significant addition to this area of research. These rely on hydrodynamic forces such as the Zweifach-Fung bifurcation law[2–4], Fahraeus effect[5], Fahraeus-Lindqvist[6], shear-thinning[7], and plasma-skimming[8]. Passive devices result in lower shear stresses and are preferred in this regard.

Recent advancements in fabricating low-cost, micron-sized rectangular channels using polymers and elastomers have led to a new family of devices that can separate RBC and platelets[9,10] from plasma. Flow analysis is essential in designing this separation technology. A continuum framework that conceptualizes blood as a suspension, where plasma is interspersed with RBC, a spherical, rigid particle, is a promising first step. This notion, referred to as the theory of interacting continua, was developed by Truesdell[11]. It depends on treating several mixture components as a unified continuum, considering their local interactions in terms of drag and lift forces due to collisions. The theory has recently been adopted in various biomedical contexts, including blood flow through microchannels[12–14]. In a simplified treatment, plasma is viewed as a Newtonian fluid, while the viscosity of blood depends on the local shear rate and the hematocrit, or the volume fraction of the RBCs. While being applicable at a larger length scale (mm or more), it is expected to fail at the length scale of individual RBCs. At these microscopic scales, the flow characteristics are greatly affected by how densely the red blood cells are packed and their flexibility. Mesoscale simulations, such as dissipative particle dynamics (DPD), can study the flow and migration of constituents at the RBC length scales. However, they incur prohibitive computational cost while lacking the scalability to larger physical domains. The continuum approach emerges as a viable alternative for investigating particle-laden blood flow in microchannels, where CFD techniques are applicable[15]. The particulate nature of blood and its influence on flow distribution are systematically revealed when the channel cross-section is comparable to the length scales of RBC.

Fahraeus[5] observed a distinctive behavior of RBCs in blood flow within microchannels of a diameter below 300 µm, with the particles migrating towards the centre. Leighton and Acrivos[16] observed a similar phenomenon in suspension experiments in a cup-and-cone rheometer and termed it "shear-induced migration". These observations were mathematically explained by Philips[17], who delineated the fluxes of particle migration and collision frequency for species transport. This approach is now referred to as the diffusive flux model (DFM). There are alternatives to DFM models, such as multiphase flow and volume of fraction simulations, which require solving the two-phase equations, but in DFM modelling, the simulations are carried out using single-phase equations, reducing the computational requirements. It addresses the shear-driven movement of particles by incorporating two essential terms into the mass transport equation for particles in a flowing liquid. The first term arises from the gradient in particle interaction frequency, while the second term results from the gradient in effective viscosity. Realistic blood flow studies can be conducted by integrating the DFM formulation within the continuum approach, coupled with a realistic rheological model for blood viscosity.

Researchers have also adopted the DFM model to study the flow of suspensions in various geometries. Chandran et al.[15] used finite volume discretization of the governing equations coupled with DFM and the viscosity model developed by Apostolidis and Beris[18] specifically for blood. This approach was critically examined in a later work of Giri et al.[19]. These simulations agreed quite well with the substantially more detailed DPD simulations conducted by prior researchers[20–22]. This close alignment indicates that the DFM methodology effectively captures the underlying dynamics of particulate transport in a liquid-particulate mixture when the channel dimensions are larger but comparable to those of the particulates. As observed in the earlier studies[18,19], the results are consistent with DPD even when channel dimensions are about five times larger than the RBCs. Additionally, this coupled framework on the continuum scale proves economical in terms of effort, time, and cost relative to detailed mesoscale simulation techniques. The gains are particularly significant when repeated simulations are required for geometrically optimizing biomedical devices involving blood flow, as in this study.

In the present work, we adopt the DFM framework to analyze a passive separator used for platelet separation from blood. The primary objective is to examine the sensitivity of the design parameters of a trifurcated microfluidic platelet separator with respect to its effectiveness. The trifurcated platelet separator device uses the shear-induced migration of RBCs within the suspension. Here, the particles migrate from a high shear region to a low shear one, increasing the particle concentration at the centre. This creates a region of low concentration in high shear zones near the wall of the microchannel. Attached to the high shear zone, separator arms extract the RBC-depleted liquid. For an effective extraction of low-concentration liquid, the separator unit is repeated in the flow direction of the platelet separator device. A comprehensive examination of geometrical parameters, flow rates, and RBC dilution is conducted within a trifurcated channel with separator arms, whose geometric details were first described in an earlier study[23,24]. The present work carries out an analysis of this device using the diffusive flux model as described in the following section.

## 2. Numerical methodology

The schematic drawing of the region of interest, physical mechanisms, and the governing equations, followed by a numerical solution of the system of coupled nonlinear partial differential equations, are described in the present section. In addition, grid independence and solver validation with the published literature are also presented.

### 2.1. Geometry and Range of Parameters

The schematic of the trifurcated channel and the relevant geometric dimensions are given in Figure 1. The lengths of the two separator arms are kept equal. The inlet width is given a symbol $H$ (= 80 µm), and the separator arm width ($H_1$) = 60 µm, while the length of the separator arm is $16H$. The total length of the trifurcated device is $L = 50H$. The angle ($\alpha$) between the main channel and separator arm and the width of the main channel ($H$) chosen for simulations are summarized in Table 1. Specifically, the separator angles selected are 45° and 90°. The corresponding inlet flow rates with the Reynolds number based on the effective blood viscosity ($\mu = 0.0035$ Pa s) are also summarized in Table 1. The geometric dimensions selected for analysis have been adapted from previous work on microfluidic separator devices[9-10, 20-22].

The length upstream of the trifurcation point is $32H$ and is chosen in such a way as to eliminate the entrance effects and provide a pathway for the concentration profile to be fully developed. Beyond this distance, the separator arms diverge from the channel to extract the platelet-enriched plasma from the zone near the device walls. After the trifurcation zone, the channel is again a straight segment where the particles migrate towards the centre. In applications, the trifurcation unit shown in Figure 1 is a part of a repeating assembly. In the present work, the focus is on flow and species concentration distributions in a single unit.

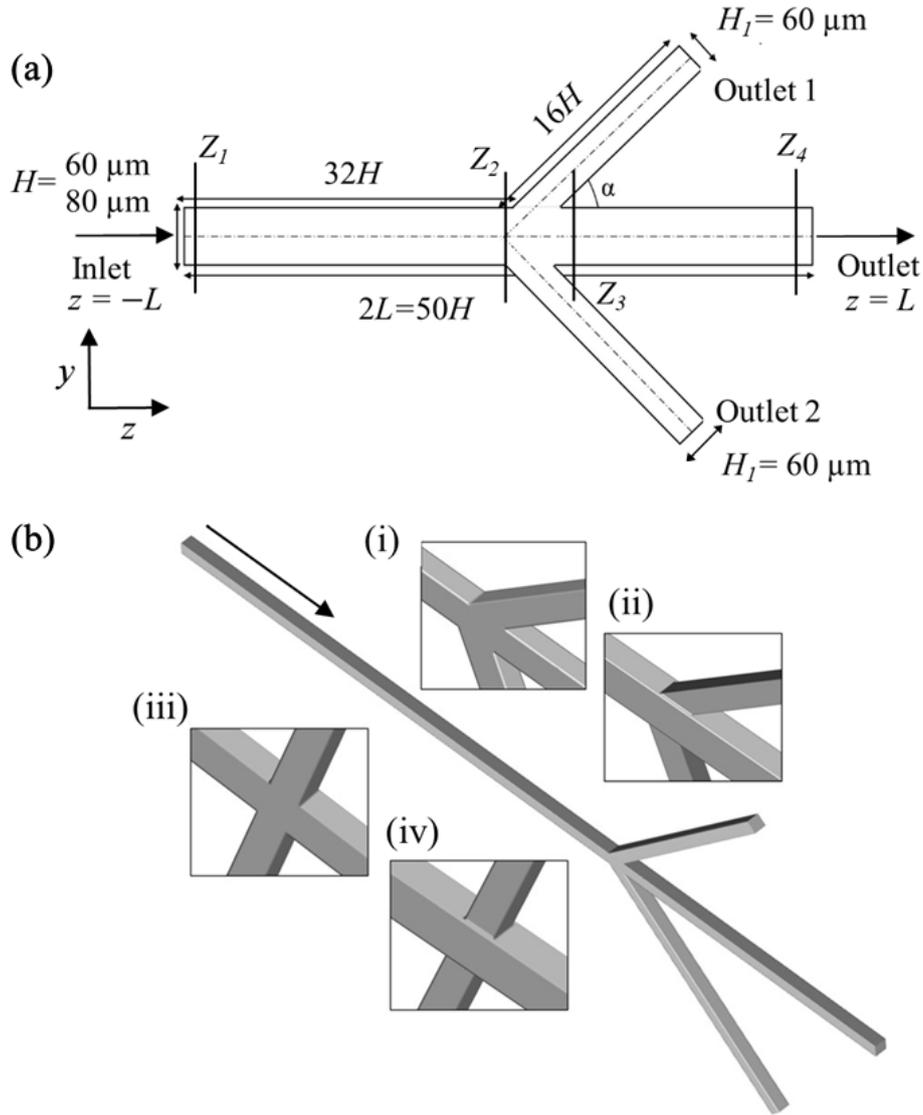

Figure 1: (a) Schematic of the separator device with selected locations highlighted for data analysis. (b) The 3D trifurcated channel is displayed along with the magnified view of separator arms at $\alpha = 45°$ (i and ii) and 90° (iii and iv) for $H = 60$ µm and 80 µm.

Table 1. Parameters considered for the analysis of hematocrit separation from plasma.

| Channel width, $H$ (µm) | Separator angle, $\alpha$ (°) | Inlet flow rate, $Q$ (ml/min) | Reynolds number (Re) for $H = 60-80$ (µm) | Average hematocrit, $\phi$ | Blood temperature, $T$ (K) |
|---|---|---|---|---|---|
| 60 | 45 | 0.25 | 11.8 - 15.77 | 0.2 | 298 |
| 80 | 90 | 0.5 | 23.65 - 31.54 | 0.3 | 310 |
| - | - | 1.0 | 47.31 - 63.09 | 0.4 | - |

## 2.2. Governing Equations

Blood is taken to be an incompressible liquid whose flow is determined by the mass balance and momentum equations[15,19]. The incompressibility constraint in the form of the continuity equation and the momentum equations for a variable viscosity fluid are solved jointly with the species transport equation in three dimensions as given below:

$$\nabla \cdot \vec{u} = 0 \tag{1}$$

$$\rho\left(\frac{\partial \vec{u}}{\partial t} + \vec{u}\cdot\nabla\vec{u}\right) = -\nabla p + \nabla \cdot \bar{\bar{\tau}} \tag{2}$$

Here, $\vec{u}$ is the velocity vector, $p$ is pressure, and $\bar{\bar{\tau}}$ is the deviatoric part of the total stress tensor. In a non-Newtonian fluid framework, it is a function of the local hematocrit as well as the strain rate, and is given as

$$\bar{\bar{\tau}} = \mu(\phi, \dot{\gamma})\dot{\bar{\bar{\gamma}}} \tag{3}$$

Here

$$\bar{\bar{\gamma}} = \frac{1}{2}\left(\nabla\vec{u} + \nabla\vec{u}^T\right) \tag{4}$$

The apparent viscosity $\mu$ is a function of the hematocrit (i.e., RBC concentration $\phi$), and the magnitude of the strain rate, which is expressed as

$$\dot{\gamma} = \sqrt{2\bar{\bar{\gamma}}:\bar{\bar{\gamma}}} \tag{5}$$

The variation of the hematocrit in the channel is determined by solving the hematocrit transport equation, which incorporates the fluxes responsible for particulate migration of RBC in blood. In the present formulation for apparent viscosity, RBCs are taken to be the dominant constituent among species (~ 40%), and the presence of platelets (< 2%) and WBCs is neglected. The species transport equation is written for RBC migration in the following form:

$$\rho\left(\frac{\partial \phi}{\partial t} + \vec{u}\cdot\nabla\phi\right) = \nabla \cdot \vec{J} \tag{6}$$

$$\text{where } \vec{J} = \vec{J}_c + \vec{J}_\mu + \vec{J}_\kappa + \vec{J}_b \tag{7}$$

Here, the RBC concentration is given a symbol $\phi$, defined as the ratio of total RBC volume to total blood volume. Hence, plasma concentration can be defined as $(1 - \phi)$. The diffusive flux $\vec{J}$ includes the sum of fluxes arising from the variation in the frequency of particle collisions, effective viscosity, and streamline curvature[25]. It also includes Brownian diffusion $\vec{J}_b$, related to gradients in RBC concentration.

Following the work of Philips[17], non-uniform particle interaction in the flow field tends to move the particle from the zone of high shear rate to one of low shear rate. This particle interaction frequency is estimated in terms of the shear rate as $\dot{\gamma}\phi$ and varies at the length scale of the order of the particle diameter (here, RBC diameter, $a$). Nonuniform collisions lead to an unsymmetric pressure variation that, at the scale of a particle, will appear as an unbalanced lift. The resulting diffusive mass flux generated by these particle interactions is written as[16-17]:

$$\vec{J}_c = k_c a^2 \phi \nabla(\phi \dot{\gamma}) \qquad (8)$$

Additionally, a variation in the hematocrit normal to the wall results in an effective viscosity gradient and hence an unbalanced drag on the particle. As a result, particles in the high viscosity gradient region move a larger distance compared to others in the low viscosity zone[25]. The mass flux due to the spatial variation in effective viscosity is given as[16-17]:

$$\vec{J}_\mu = k_\mu a^2 \phi^2 \left(\frac{\dot{\gamma}}{\mu_{app}}\right) \nabla \mu_{app} \qquad (9)$$

Following the work of Krishnan et al.,[26], the particles also tend to migrate towards the regions of lower streamline curvature where centrifugal forces are smaller. The mass flux arising from the differences in streamline curvatures of the flow path is given as:

$$\vec{J}_\kappa = k_\kappa a^2 \phi^2 \dot{\gamma} \nabla \ln R \qquad (10)$$

In Equations 8-10, $k_c$ (= 0.41), $k_\mu$ (= 0.64), and $k_\kappa$ are experimentally determined parameters, $r$ is the radius of the curvature of local streamlines, and $R = a/r$ denotes the streamline curvature relative to the mean curvature of the particle. Since the trifurcated channel of the present study is mostly straight, particulate flux owing to streamline curvature may be neglected.

With $D_{RBC}$ as the RBC diffusivity in the suspension, the flux induced by Brownian motion is given as:

$$\vec{J}_b = D_{RBC} \nabla \phi \qquad (11)$$

For large particulates such as RBCs, diffusivity ($D_{RBC}$) is of the order O($10^{-9}$) m$^2$/s the contribution of Equation 11 to Equation 6 is expected to be negligible.

The diffusive flux model is coupled with the incompressible form of continuity and momentum equations in terms of the apparent viscosity, which, in turn depends on the local hematocrit concentration. This functionality for human blood is discussed further in Section 2.4.

## 2.3. Boundary Conditions

The following initial and boundary conditions are applied to solve the three-dimensional form of the unsteady governing equations in a trifurcated channel.
1. The initial state is the rest condition of the liquid suspension.
2. Steady uniform inflow velocity is prescribed at the inlet in terms of an equivalent flow rate specified in Table 1. A uniform average hematocrit $\phi = 0.4$ is prescribed on the inflow plane of the microchannel.
3. At the three outlets of the microchannel, the gradient outflow flow condition is selected for velocity and hematocrit distribution, respectively, whereas zero-gauge pressure is enforced.
4. A no-slip boundary condition is applied for the velocity components on the solid walls.
5. For hematocrit, the no mass flux condition $\vec{n} \cdot \vec{J} = 0$ is enforced at the channel walls.

## 2.4. Blood rheology model

The rheology model should be correctly selected for determining the apparent viscosity of blood. In addition to shear rate, the model parameters additionally depend on the local hematocrit, whose distribution is determined by the species transport equation. Starting with the Casson viscosity model, Apostolidis and Beris[18] developed a model specifically for blood by including its microscale constituents. The resulting viscosity model, derived from a vast body of rheometer data is summarized below as follows. The apparent viscosity at any location and time instant is obtained as

$$\mu_{app} = \left( \sqrt{\frac{\tau_0}{\dot{\gamma}}} + \sqrt{\mu_c} \right)^2 \tag{12}$$

Here, $\dot{\gamma}$ is the magnitude of the strain rate tensor, Equation (5). Apostolidis and Beris proposed that blood has a yield stress that depends on the hematocrit concentration, given as follows:

$$\tau_0 = \begin{cases} (\phi - \phi_c)^2 [0.508 c_f + 0.4517]^2, & \text{if } \phi > \phi_c; \\ 0, & \text{if } \phi < \phi_c; \end{cases} \quad (13)$$

The threshold value of hematocrit beyond which the yield stress is non-zero depends on the fibrinogen concentration, $c_f$, which serves to bind the RBCs, and is given as

$$\phi_c = \begin{cases} 0.312 c_f^2 - 0.468 c_f + 0.1764, & c_f < 0.75; \\ 0.0012, & c_f \geq 0.75; \end{cases} \quad (14)$$

An intermediate quantity, referred as Casson's viscosity $\mu_c$ is specified as a function of the hematocrit concentration with an Arrhenius temperature dependence:

$$\mu_c = \mu_p \left(1 + 2.0703\phi + 3.722\phi^2\right) \exp\left\{-7.0276\left(1 - \frac{T_0}{T}\right)\right\} \quad (15)$$

For fully developed conditions in the channel, the local shear rate at the centreline is $\dot{\gamma} = 0$, which makes the $\vec{J}_\mu$ flux zero (Equation 9), resulting in the particle migrating to the centre without any opposing flux. This will cause an apparent viscosity approaching infinity, leading to numerical instability during simulations. To ensure the stability of the solver at fully developed conditions, the shear rate is regularised as proposed by Miller and Morris[27], i.e.

$$\dot{\gamma} = \dot{\gamma} + \dot{\gamma}_r \quad (16)$$

where the non-local shear rate $\dot{\gamma}_r \ll \dot{\gamma}$ and given as $\dot{\gamma}_r = \frac{\varepsilon U_{max}}{L_c}$. Here, $U_{max}$ is the maximum velocity in the channel based on the Newtonian viscosity, and $\varepsilon$ is the ratio of the particle radius and the half-channel width of the trifurcated microchannel.

Viscosity related parameters of blood are summarized in Table 2 for a temperature of 310 K. A parametric study with respect to this temperature is discussed in Section 3.4.

Table 2: Parameters considered for the blood viscosity model at the temperature of 37°C .

| Blood Parameters | |
| --- | --- |
| RBC diameter, $a$ | 8.2 μm |
| Blood sample temperature, $T$ | 310 K |
| Reference Temperature, $T_0$ | 296.1 K |

| Plasma viscosity, $\mu_p$ | $1.67 \times 10^{-3}$ N s /m$^2$ |
|---|---|
| Blood density, $\rho$ | 1060 kg/m$^3$ |
| Fibrinogen concentration, $c_f$ | 0.125 g/dl |

## 2.5. Numerical Solution and Grid Independence

The migration of RBCs in the trifurcated channel is simulated using the coupled system of mass, momentum, and species transport equations. The species equation incorporates diffusive fluxes, while the apparent viscosity is determined using the Apostolidis and Beris model[18]. The properties of blood are summarized in Table 2. The coupled system is solved by developing a finite volume solver within an open-source C++ toolbox (Open Field Operation and Manipulation – OpenFOAM-v10) used for developing customised differential equation solvers, integrating the respective modules to create a single 3D unsteady solver. A transient solver - pimpleFoam, based on the PISO+SIMPLE algorithm, is used to develop the solver for mass and momentum equations (1 and 2) coupled with the one for hematocrit transport (Equation 6). The present study is focused on steady-state conditions. Hence, the unsteady form of the governing equations is integrated in time till a steady state is reached. For the geometry and flow properties considered, a steady state was reached within a physical time scale of one second.

The developed OpenFOAM solver is executed on the PARAM SANGANAK supercomputing facility at IIT Kanpur (India). The solver is parallelized using the OpenMPI library, enabling efficient domain decomposition and communication between processors. Each simulation is executed on 48 cores to ensure adequate speed-up.

For the grid convergence test, four individual meshes labelled M1, M2, M3, and M4 are selected with an increasing number of elements (Figure 2). The results shown here are for 3D hexahedron mesh elements (Figures 2(b-c)) that yield superior mesh quality for rectangular cross-sections. The governing equations have been solved for a steady uniform inlet flow rate, and the average velocity at the main channel outlet is monitored. It is observed that the differences in predictions of the outlet flow rate between the M3 and M4 meshes are minor, Figure 2a; hence, the M3 mesh has been adopted for further analysis of the separator device.

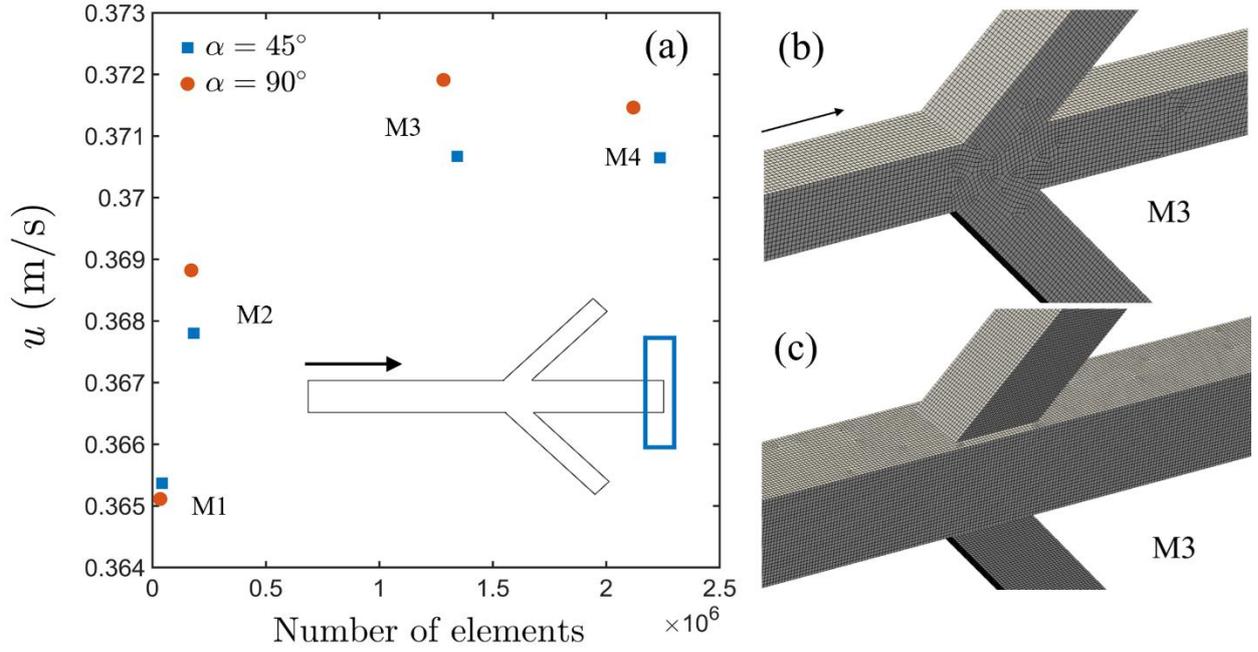

Figure 2: (a) Average velocity at the device outlet is shown for the selected meshes with 0.25 ml/min inlet flow rate (inlet velocity of 0.651 m/s, $H$=60 µm) for both separator angles. The meshes use hexahedron elements for better quality, as shown in (b) and (c) for the M3 mesh for $H$ = 60 µm and 80 µm, and a separator arm angle of 45°.

## 2.6. Solver Validation

The flow solver developed in OpenFOAM, incorporating the diffusive flux model, is validated against the experimental data available in the literature. The viscosity model used for the validation is that of Krieger[28], where apparent viscosity of the suspension is given as

$$\mu_{app} = \mu_0 \left(1 - \frac{\phi}{\phi_{max}}\right)^{-1.82} \quad (17)$$

Here, $\phi_{max}$ is the maximum volume fraction occupied by the particles in suspension at the axis of the channel, and $\mu_0$ is the viscosity of the background liquid. When the local hematocrit $\phi$ reaches its maximum value, i.e., $\phi_{max}$, the local viscosity will increase to infinity, resulting in the divergence of the solver. Therefore, a threshold value of $\phi$ is chosen near the channel axis as $\phi_\delta = \phi_{max} - \delta$ ($\delta \sim 10^{-2}$).

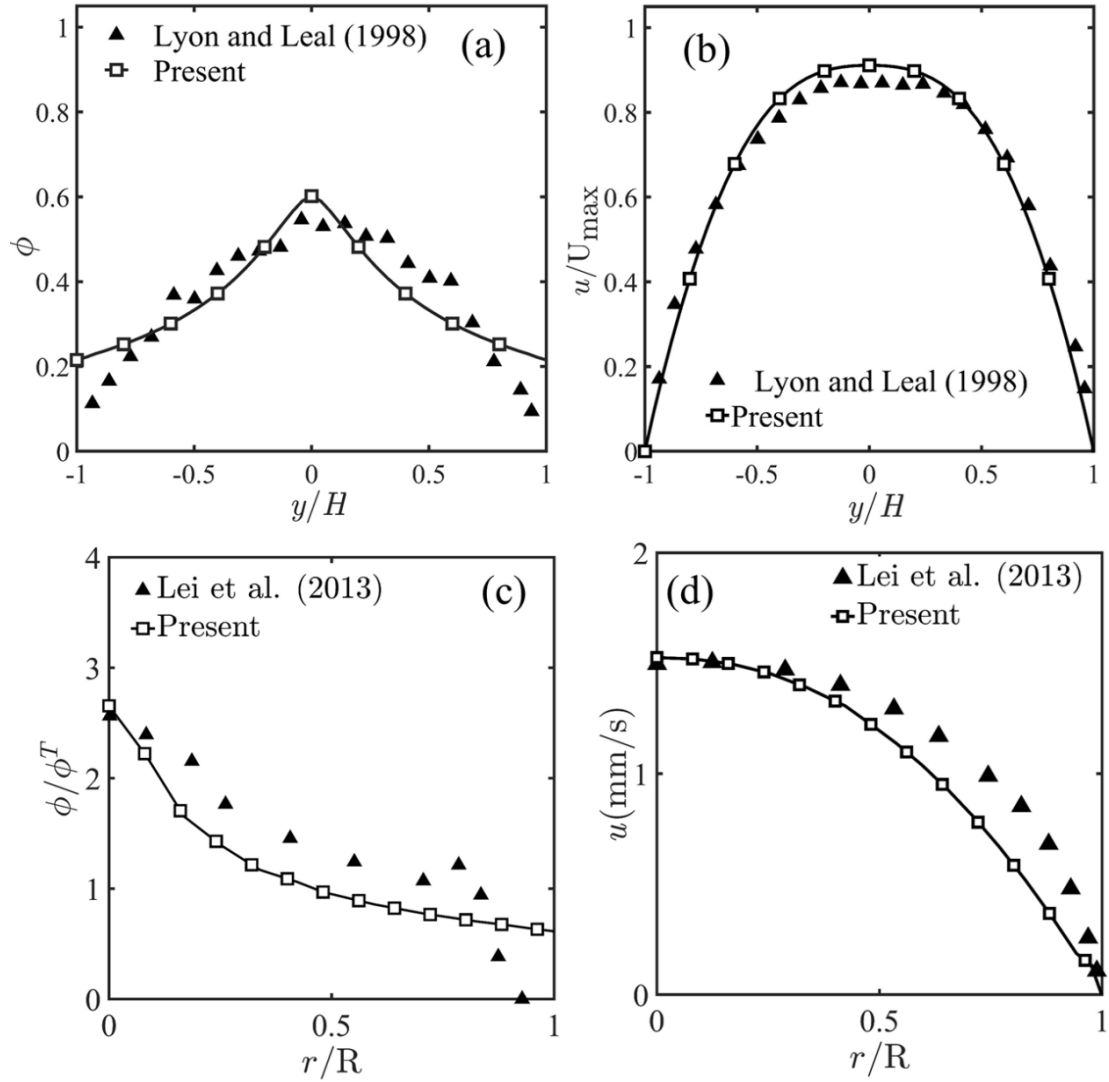

Figure 3: (a) Concentration distribution in a square channel is shown along with the experimental results of Lyon and Leal[29]. (b) The simulated velocity distribution is also compared with experiments. (c) Species concentration and (d) axial velocity profiles are compared with the DPD study of Lei et al.[20] for a circular conduit with average hematocrit $\phi = 0.3$.

In Figures 3(a) and 3(b), the concentration and velocity profile data obtained from the rectangular channel simulation are compared with the experimental data of Lyon and Leal[29], respectively. In a later study, Lei et al[20] presented the concentration profiles from a DPD study for blood flow in a circular conduit. In Figures 3(a-d), the resultant distribution of velocity and hematocrit from the present study is compared with the experimental work and the DPD simulations, and a good agreement is seen. An expected discrepancy is noticed near the wall region in the comparison with DPD simulations (Figures 3(c-d)). This is due to the limitation of the continuum model in capturing wall interactions stemming from the finite size of red blood cells (RBCs). In dissipative particle dynamics (DPD) simulations, the location of an RBC is represented by its centre of mass. Therefore, RBCs observed near the wall appear slightly away from the wall by a distance approximately equal to the major axis of the RBC.

## 3. Results and Discussion

The trifurcated microchannel behaves as a platelet separator in the following manner. Shear-induced migration displaces the dominant particles, namely RBCs, toward the axis while viscosity gradient transports them in the opposite direction, toward the wall. The net result is the enrichment of blood with hematocrit near the axis and that of the platelets near the wall. These trends are reproduced in the simulation, where it is observed that the hematocrit concentration $\phi$ increases near the central axis. The plasma, enriched in platelets, moves closer to the wall by a process known as margination. The liquid taken out from the separator arms on the sides of the trifurcated channel is rich in platelets. In the present section, the velocity distribution of the mixture and RBC concentration in the trifurcated separator device is presented over a range of parameters. The performance of the device is gauged in terms of the cell-free layer thickness in the main channel and the overall separator effectiveness. The origin of RBC separation in plasma is discussed in terms of the migration fluxes in the microchannel. The sensitivity of the device performance to bulk temperature and numerical considerations, such as 2D versus 3D simulations, and coupled versus decoupled models, are examined.

### 3.1. Velocity and Hematocrit distribution

Color contours of the fluid velocity magnitude for the selected inlet flow rate are shown in Figure 4 for a separator angle of 45º and Figure 5 for 90° in a separator device with $H=60$ µm. To quantify the velocity distribution further, these are shown over the selected planes $Z_1$-$Z_4$ referred to in Figure 1. Similarly, the hematocrit concentration is shown as contour plots and line plots in Figures 6 and 7 for $H = 60$ µm with separator angle $\alpha = 45°$ and 90°. The velocity distribution is shown as a function of the normalised coordinate $y$ across the flow over selected positions of the microchannel. The velocity profile attains a fully developed shape upstream of the trifurcation section.

Post-trifurcation, the average velocity in the main channel decreases nearly by a factor of three when compared to a location upstream of the trifurcation junction. A similar trend in velocity distribution is also observed for the 90° separation angle across the three inlet flow rates. This observation of equal division among the three branches is expected since the separation arm in Figures 4-7 is of the same cross-section as the main channel (i.e., $H_1 = H = 60$ µm).

In Figures 6 and 7, the hematocrit profiles are shown for the three flow rates and two separator angles. Post-trifurcation, it is seen that the hematocrit amount in the straight section has increased. This behaviour is expected as the RBC-depleted stream that is plasma-enriched is extracted from the separator arms, ensuring a higher hematocrit concentration in the central channel beyond the trifurcation point. The effects of the separator configuration on velocity as a function of the axial coordinate near the walls and at the centreline of the channel are shown in Figure 8 for a separator angle of 45º. Differences in the velocity profile, particularly in the central part of the channel with the 90° separator, were found to be small and are not repeated. Given the symmetry in geometry, the velocity distribution remains comparable across three different flow rates for both angles, suggesting the angle parameter is less significant in the separation of the platelet-enriched plasma. The flow field distributions are crucial in determining the hematocrit distribution due to their dependence on shear-induced and viscosity-related migration fluxes. These fluxes depend on the local strain rate, influencing the hematocrit concentration and its gradients. In this respect, the mass-momentum-species transport equations are tightly coupled.

The contours of the hematocrit distribution ($\phi$) are illustrated in Figures 6 and 7, along with the line plot of $\phi$ with respect to the normalized *y*-coordinate across the flow direction. Four locations, i.e., near the inlet, before trifurcation, after trifurcation, and at the outlet, as shown in Figure 1, are studied. The figure reveals a high concentration of RBCs in the central zone in the first segment of the device, attributable to shear-induced migration away from the walls. As the channel trifurcates, the strain rates vary due to the orientation of the separators, leading to a broader concentration profile just downstream of the trifurcation. Post-trifurcation, the concentration remains high in the straight segment of the channel. Hematocrit profiles continue to develop till the outlet, retaining a higher concentration at the centre.

The flow and species concentration distributions described above are consistent across various flow rates and separator angles. As required by design, the separator branches are depleted of RBCs and carry the medically useful platelet-rich plasma. The shape and behaviour of the concentration profile provide insights into the effect of hemodynamics on particle migration within trifurcated microchannels, over and above changes arising from the channel geometry.

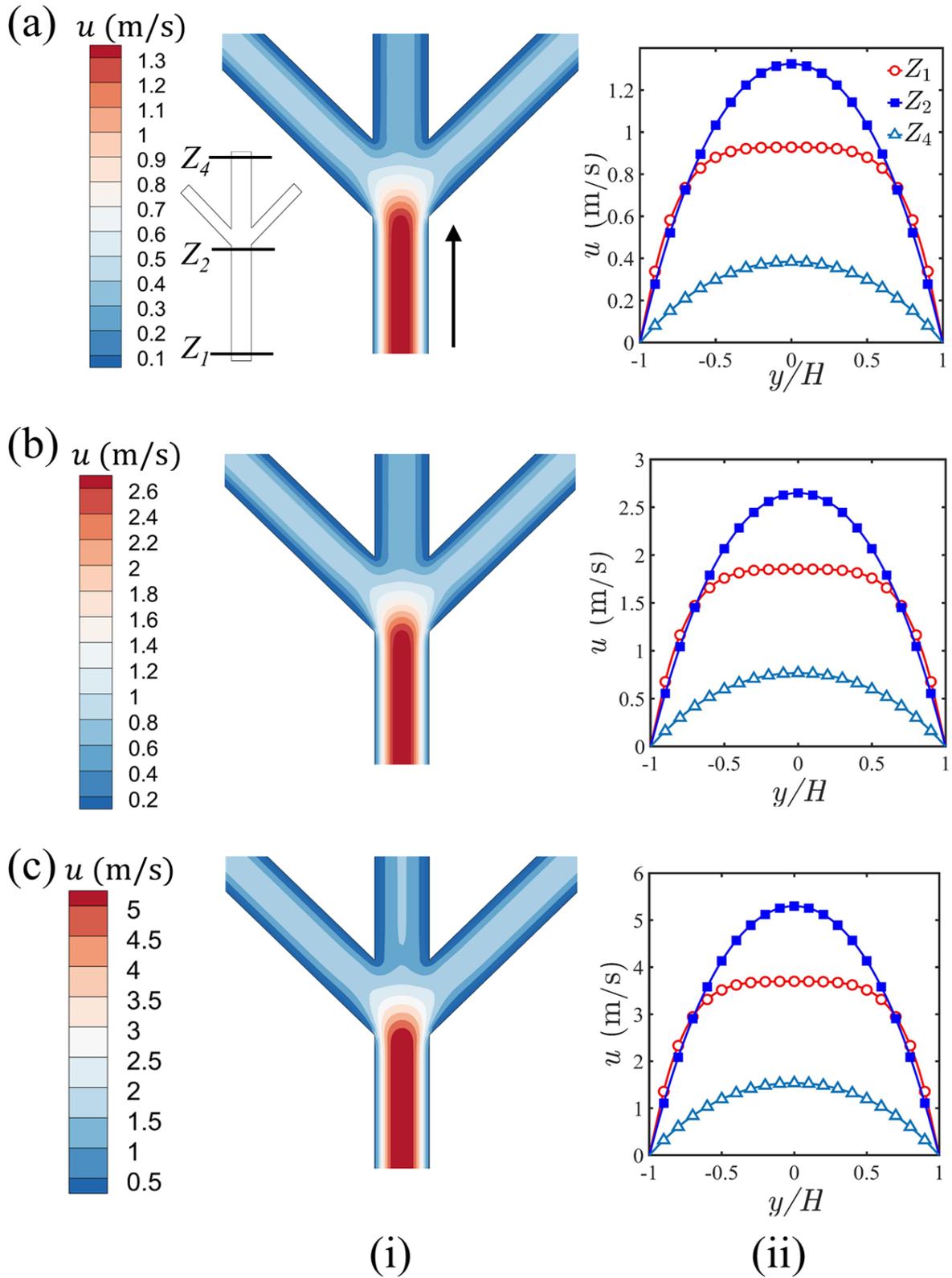

Figure 4: (i) Midplane velocity magnitude contours for $H= 60$ μm at a flow rate (a) = 0.25 ml/min, (b) 0.5 ml/min, and (c) 1 ml/min for a separator angle $\alpha = 45°$ with velocity profiles at the selected planes shown in (ii). Inlet hematocrit concentration is $\phi=0.4$.

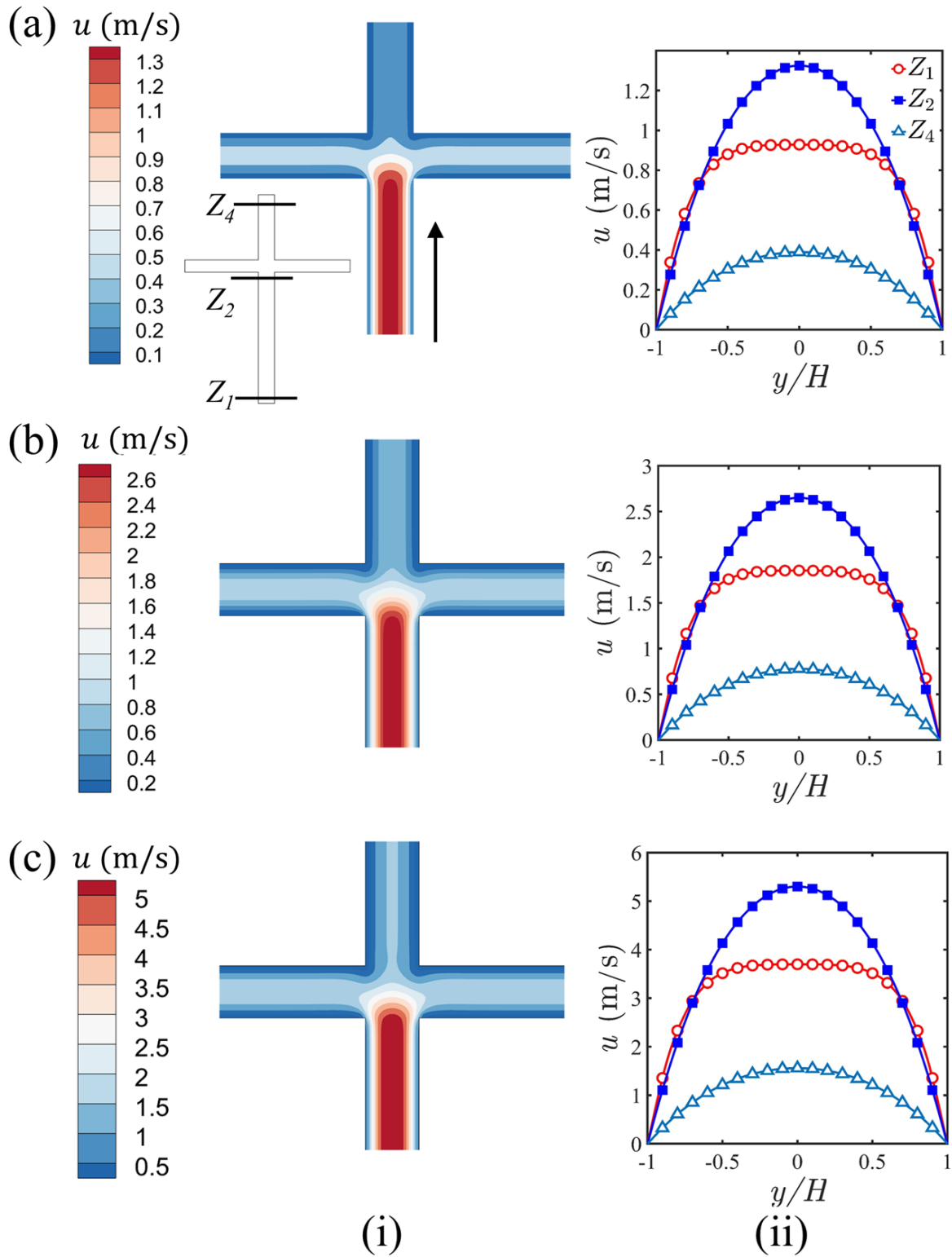

Figure 5: (i) Midplane velocity magnitude contours for $H = 60$ µm at a flow rate (a) = 0.25 ml/min, (b) 0.5 ml/min, and (c) 1 ml/min for a separator angle $\alpha = 90°$ with velocity profiles at the selected planes shown in (ii). Inlet hematocrit concentration is $\phi=0.4$.

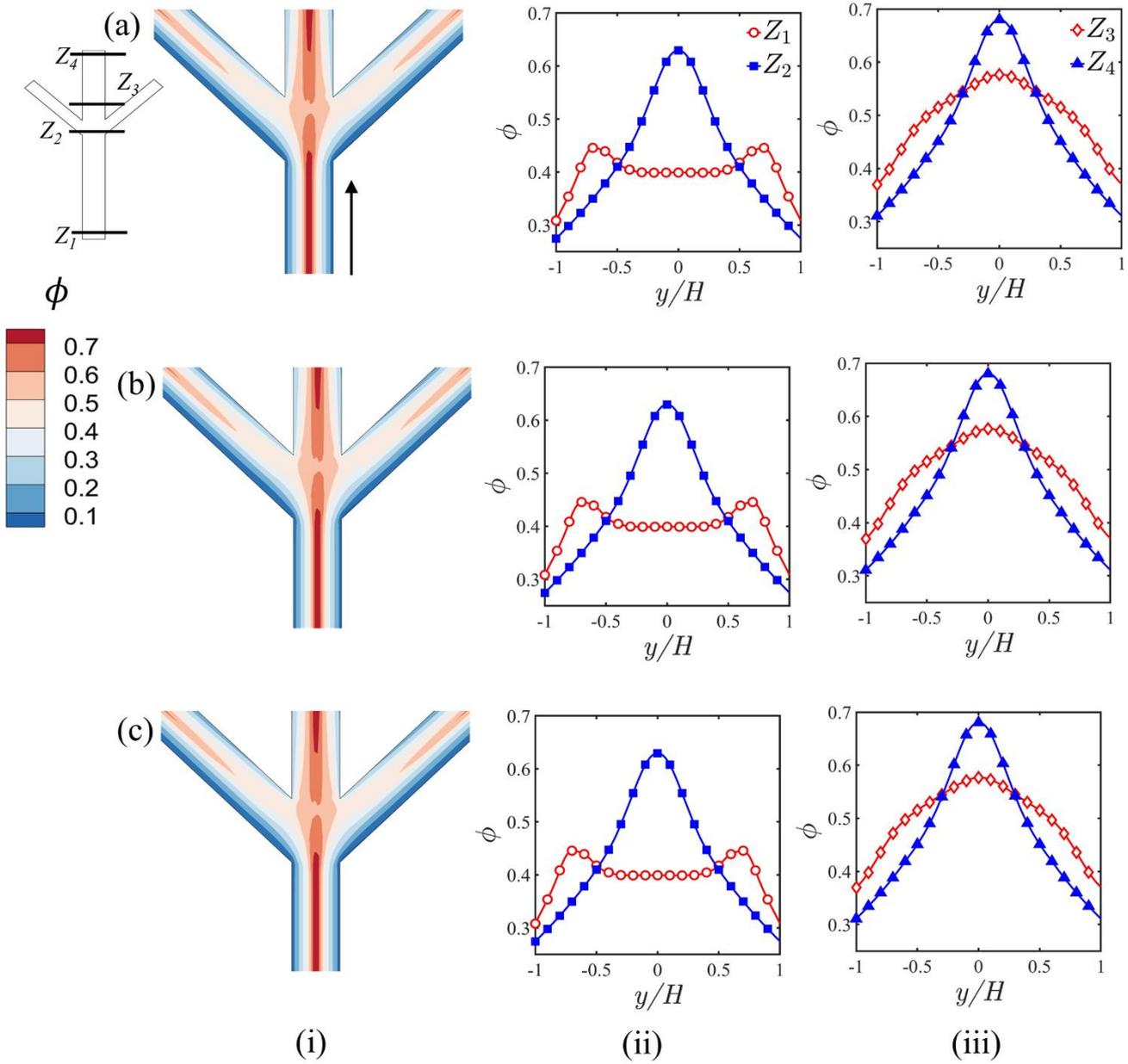

Figure 6: (i) Midplane hematocrit contours and profiles at the locations (ii) $Z_1$ and $Z_2$ and (iii) $Z_3$ and $Z_4$ of the microchannel with $H = 60$ µm. Flow rates studied are (a) = 0.25 ml/min, (b) 0.5 ml/min, and (c) 1 ml/min for a separator angle $\alpha = 45°$, with velocity profiles shown at the selected zones. Inlet hematocrit concentration is $\phi=0.4$.

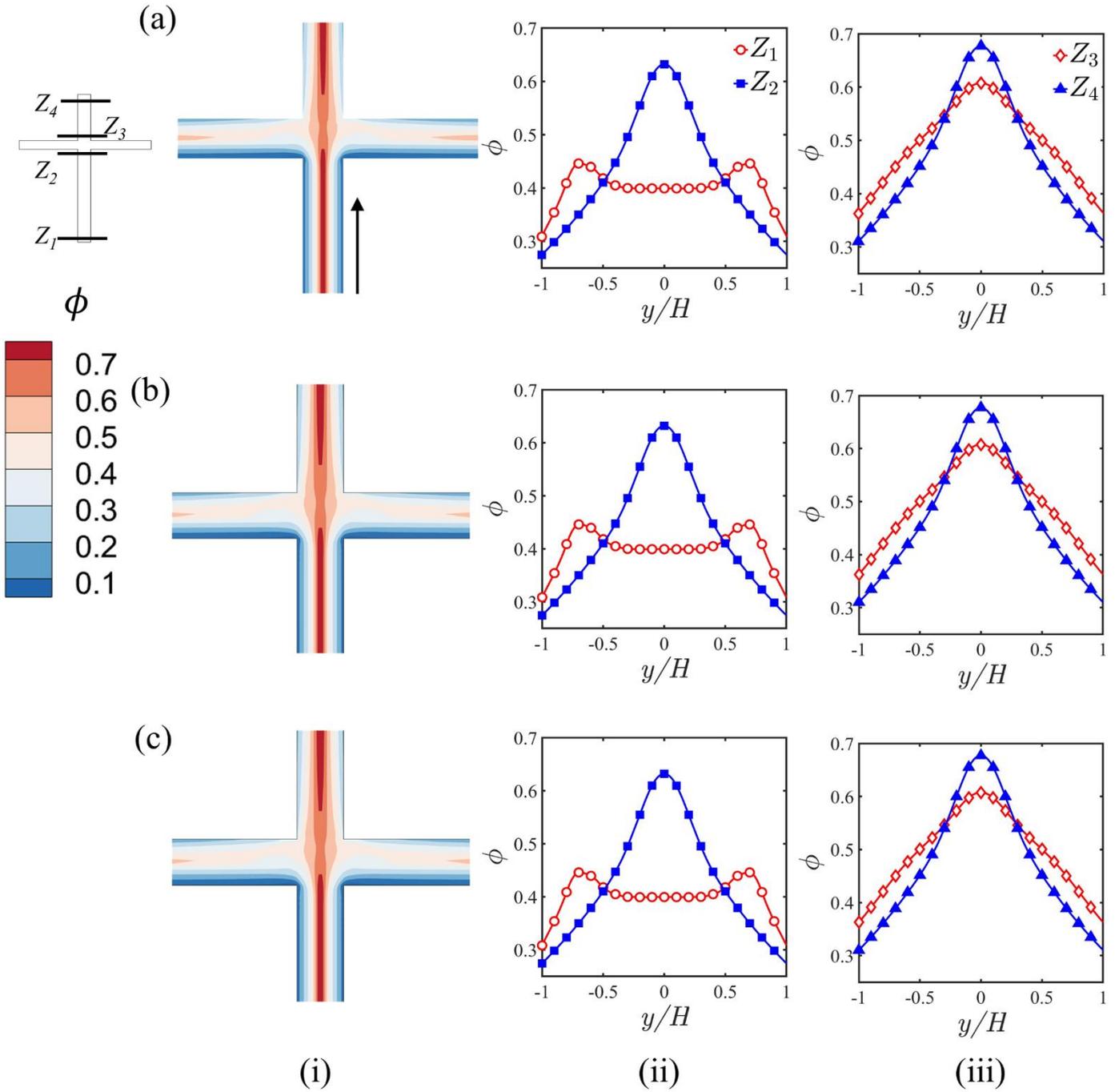

Figure 7: (i) Midplane hematocrit contours and profiles at the locations (ii) $Z_1$ and $Z_2$ and (iii) $Z_3$ and $Z_4$ of the microchannel with $H = 60$ µm. Flow rates studied are (a) = 0.25 ml/min, (b) 0.5 ml/min, and (c) 1 ml/min for a separator angle $α = 90°$ are shown along with the velocity profile at the selected zones. Inlet hematocrit concentration is $\phi=0.4$.

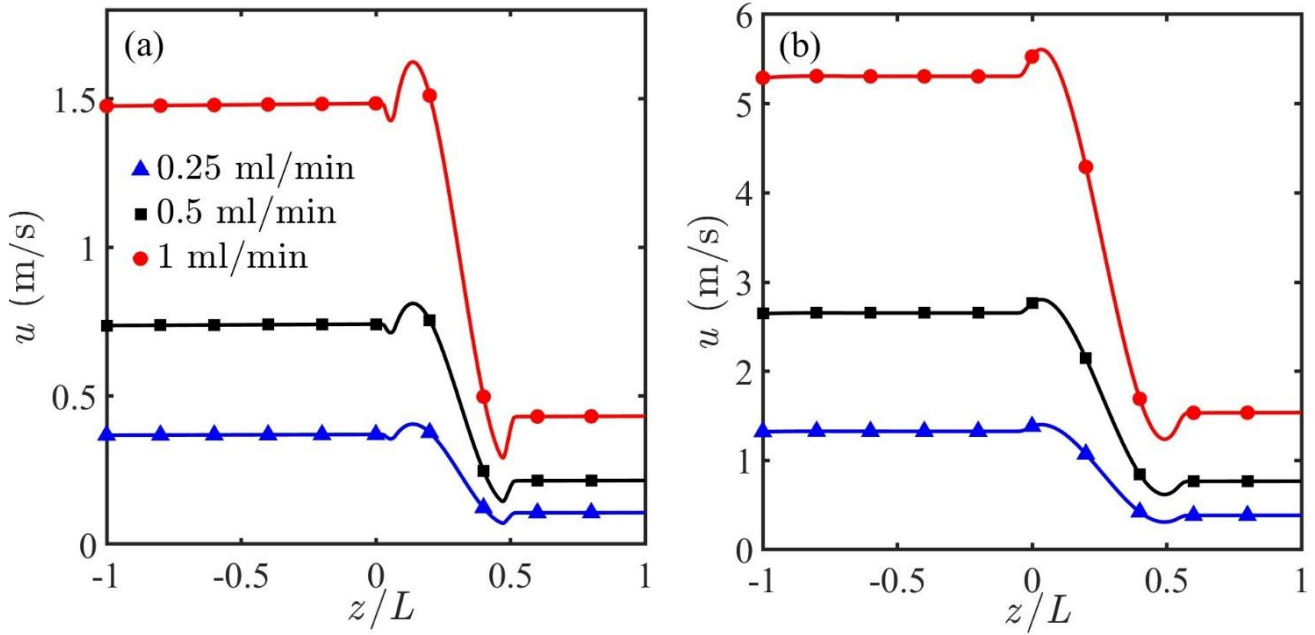

Figure 8: Midplane velocity magnitude distribution along the length of the trifurcated microchannel with $H = 60$ µm (a) near the wall and (b) at the centreline for a 45° separator angle. Similar trends are observed for the 90° separator arm of the same width and are not repeated.

The effect of the main channel width is now considered. Contours of the velocity magnitude and RBC concentration (hematocrit) are shown in Figure 9 for the microchannel with a main channel width $H = 80$ µm and separator arms of $H_1 = 60$ µm. Similar trends are observed for this microchannel in comparison to the one with $H = 60$ µm. It is also observed that the large central channel has a higher concentration, whereas the separator arm, which has less cross-sectional area, has less RBC concentration, fulfilling the Zweifach-Fung bifurcation law[3,4].

A detailed comparison between the two microchannels is provided in Figure 10 in terms of the local hematocrit and main channel width $H$. Figure 10(i) shows variations near the inlet of the main channel. Figure 10(ii) shows the profile at the outlet of the microchannel, while Figure 10(iii) shows profiles along the centreline. It is observed that a smaller channel opening increases transverse gradients and increases particle migration, thus showing higher peaks in RBC concentration. Clearly, a smaller channel opening is desirable for higher separator effectiveness.

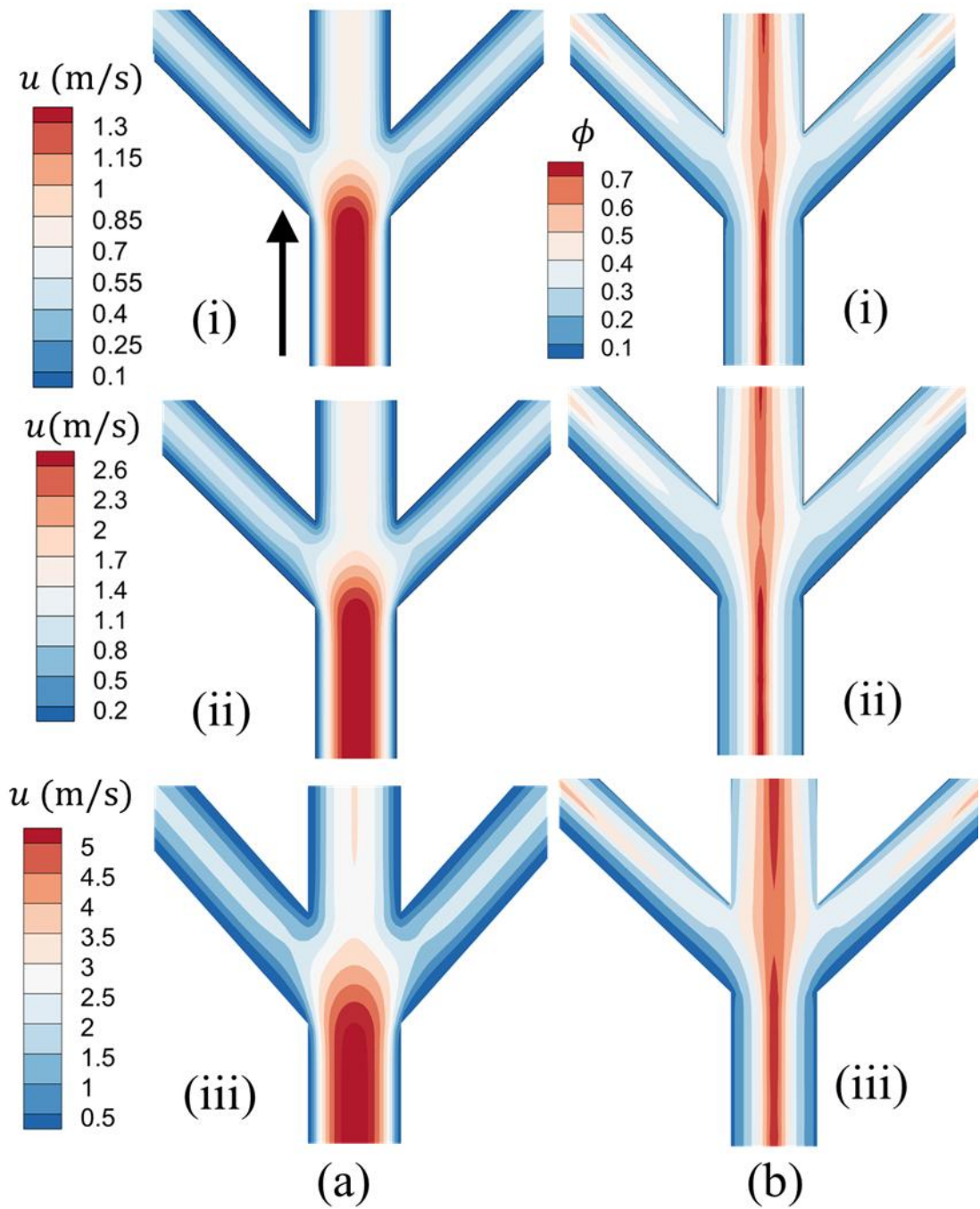

Figure 9: Midplane (a) velocity magnitude and (b) hematocrit contours for $H = 80$ µm for the flow rate $Q$ (i) = 0.25 ml/min, (ii) 0.5 ml/min, and (iii) 1 ml/min for a separator angle $\alpha = 45°$.

.

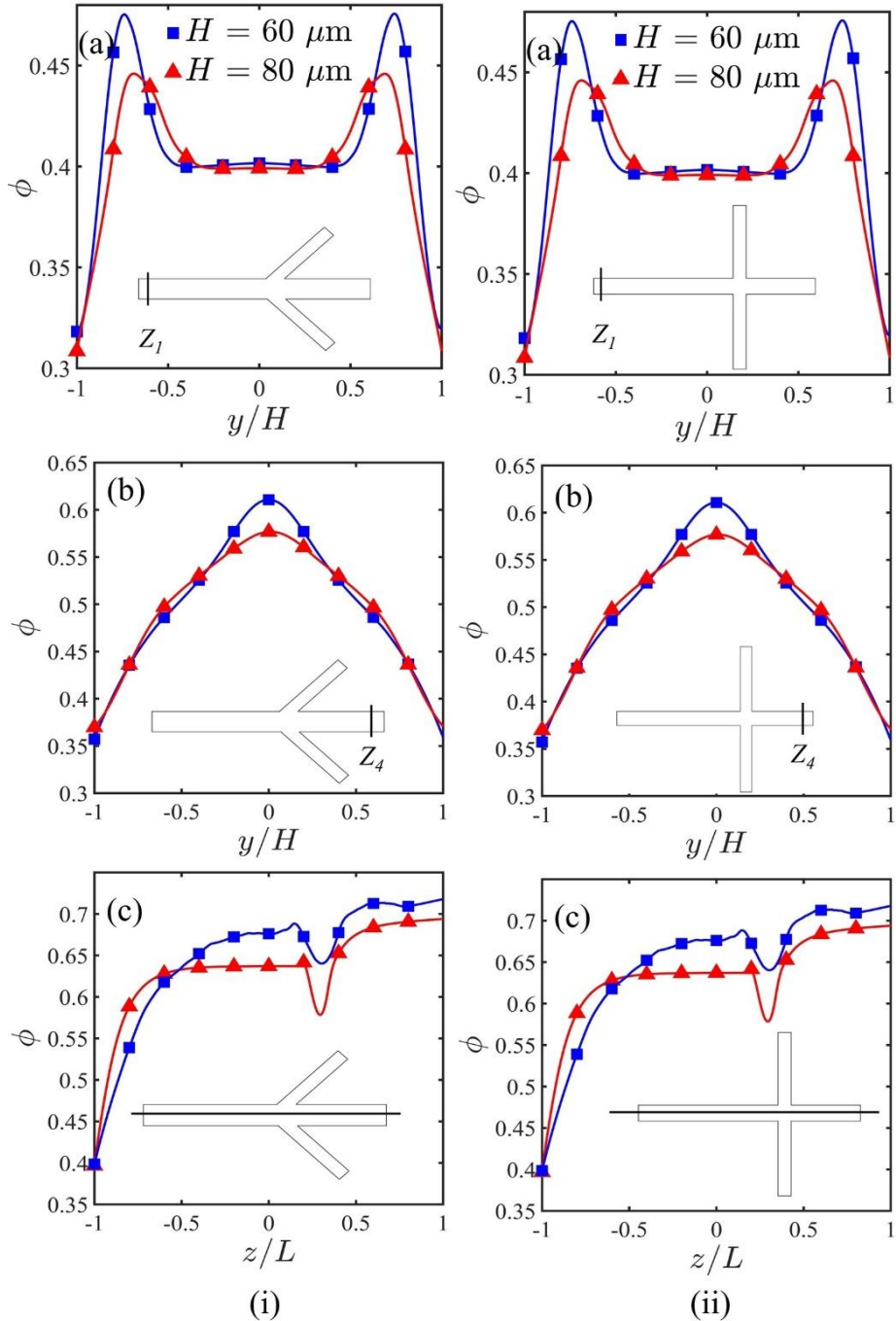

Figure 10: Hematocrit distribution near the (a) inlet, (b) downstream of the trifurcation, and (c) at the central axis at midplane of the microchannel (c) for $H = 60$ µm and 80 µm trifurcated microchannels, for the separator angle (i) $\alpha = 45°$ and (ii) 90° is shown.

In Figures 11 and 12, the distributions of hematocrit concentration and midplane velocity profiles are shown at the inlet of the separator arm, providing a comprehensive view of hematocrit distribution across the trifurcation zone. Figure 11 is focused on the velocity and hematocrit patterns for separator angles of 45° and 90°, with the main channel height fixed at 60 µm. The shown profiles illustrate the effects of increasing angular orientation of the separator arm on the separation dynamics, demonstrating distinct shifts in velocity and hematocrit layering near its inlet and outlet. These differences indicate how separator geometry can tactically be used to control the separation of red blood cells (RBC) and platelets in microfluidic trifurcations. Overall, significant differences in the velocity profiles are observed when the separator angle is increased from 45 to 90°, which alter the conveying velocity of the fluid carrying RBCs into the separator arms. The impact on hematocrit distribution is smaller in comparison.

Figure 12 examines the impact of the choice of the main channel height on hematocrit and velocity distribution, with the separator angle ($\alpha$) held constant at 45°. The analysis compares two configurations, with main channel widths ($H$) of 60 µm and 80 µm. A comparative assessment of Figures 11 and 12 reveals that the main channel width exerts a more pronounced influence on flow and hematocrit distribution in comparison to the separator angle. As illustrated in Figure 12, peak hematocrit concentration is higher in the 60 µm wide channel when compared to the 80 µm configuration. In contrast, the peak velocity magnitude is greater in the 80 µm channel. These observations further support the viewpoint that channel height is the dominant parameter in comparison to orientations of the separator arm in governing the flow characteristics and cellular separation efficiency within the microfluidic environment.

The results presented in Figures 11 and 12 emphasize the pivotal influence of geometry, specifically, the main channel height, on the modulation of flow dynamics and cell distribution within microfluidic systems. Furthermore, the study illustrates that small alterations in geometric configuration can result in substantial variations in flow behaviour and RBC-platelet partitioning. This underscores the importance of geometry-driven parametric optimization for enhancing the efficiency and functional reliability of biomedical microdevices operating under physiologically relevant conditions.

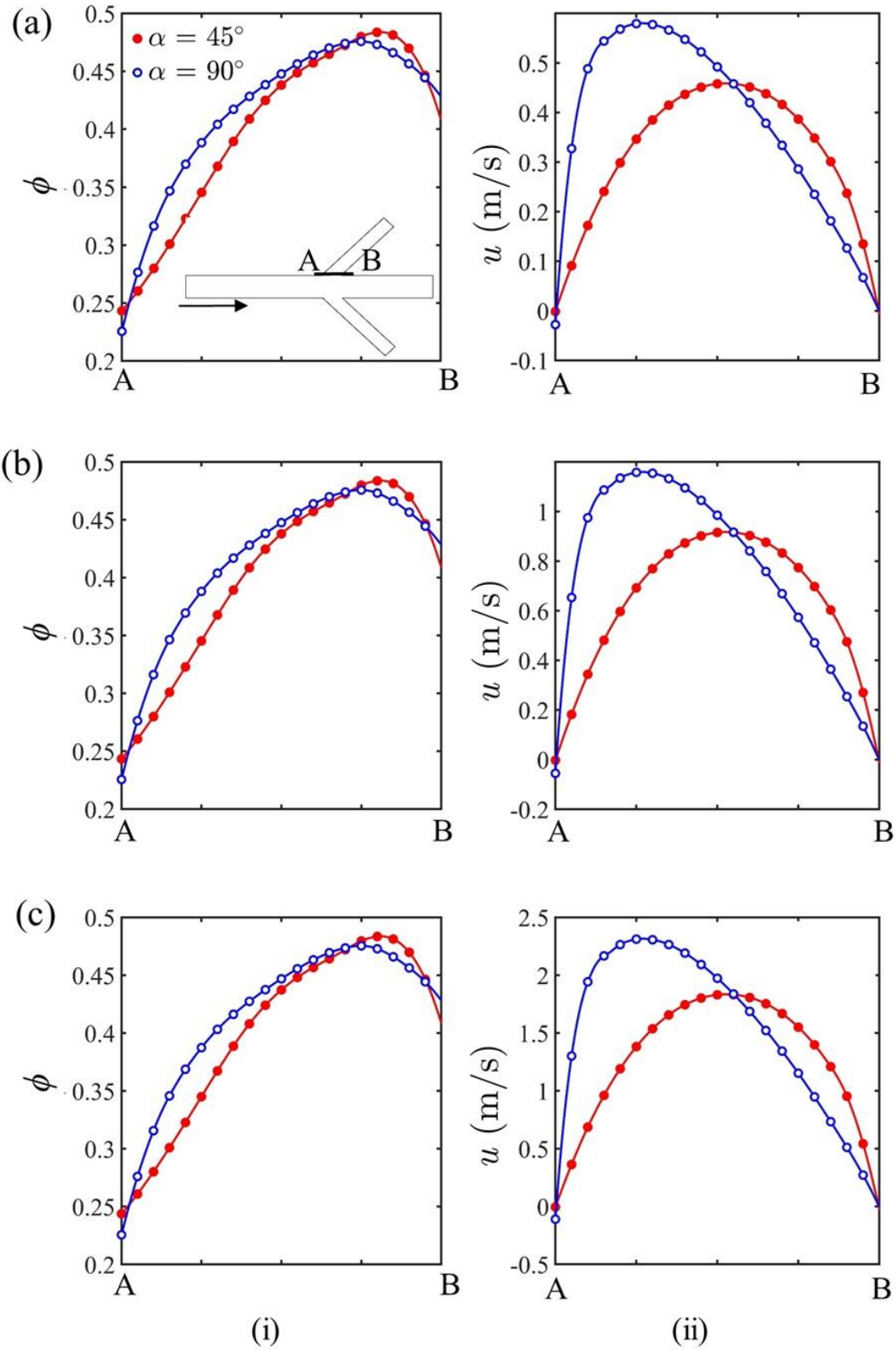

Figure 11: Hematocrit and velocity magnitude distribution over the cross-section A-B for flow rate = (a) 0.25 ml/min, (b) 0.5 ml/min, and (c) 1.0 ml/min for separator angles of 45° and 90° in a microchannel of $H = 60$ µm

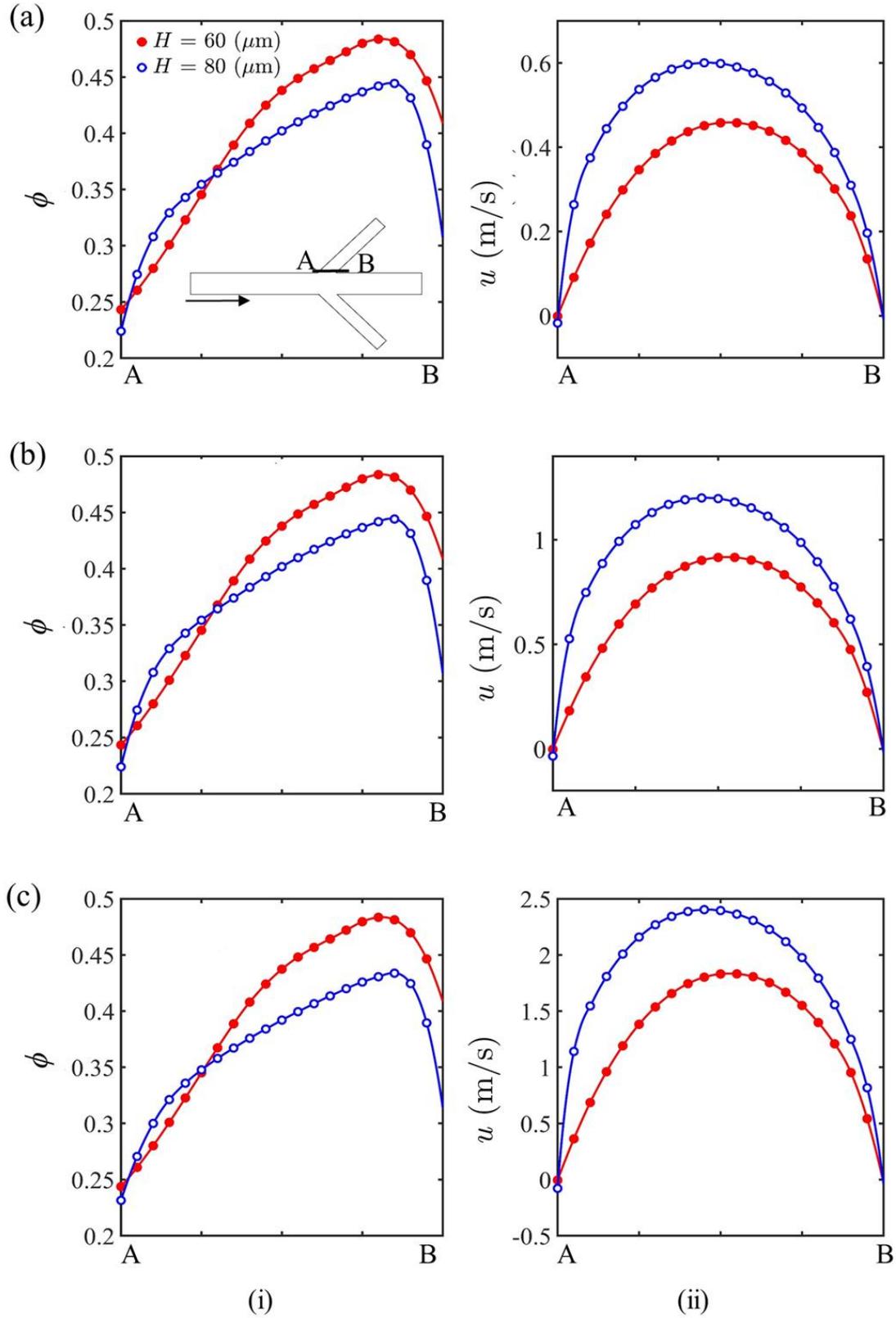

Figure 12: Hematocrit and velocity magnitude distribution over the cross-section A-B for flow rates = (a) 0.25 ml/min, (b) 0.5 ml/min, and (c) 1.0 ml/min for channel widths of 60 and 80 µm and the separator arm of angle 45°.

## 3.2. Effect of geometry on the diffusive fluxes

The migration of RBCs toward the axis of the channel is an important phenomenon exploited in the device design to concentrate RBCs in the centre and extract useful platelet-enriched plasma from the separator arms. The fluxes responsible for this migration are quantified in the present discussion. These include the flux induced by particle collisions in a field of variable shear rate $\vec{J}_c$, flux generated from the gradient in apparent viscosity $\vec{J}_\mu$, and the flux related to Brownian diffusion $\vec{J}_b$. Their respective definitions are presented in Section 2.2. The flux arising from streamline curvature is neglected in the present discussion since the device comprises straight rectangular segments. Similarly, Brownian diffusion is also negligible when compared with the collision and varying viscosity fluxes. The definitions of mass fluxes show that they are strong functions of the hematocrit concentration $\phi$, strain rate magnitude $\dot{\gamma}$, apparent viscosity $\mu_{app}$, and their gradients. The magnitudes of the three fluxes are presented as contour plots in Figure 13, depicting their behaviour in the microchannel with $H = 60$ µm and at an angle of $\alpha = 45°$ for the separator arm with $Q = 0.25$ ml/min and inlet hematocrit $\phi = 0.4$. It may be noted that the collision flux and the viscosity gradient flux are oppositely oriented in the initial segment of the separator, being away from and toward the wall, respectively.

In Figure 13, the contour of the total flux magnitude is first shown for the entire trifurcated microchannel. A magnified view of a section of the trifurcation zone around the junction is then presented. The individual magnitudes of the collision flux (Figure 13(a)) and flux induced by viscosity gradients (Figure 13(b)) are significantly higher than the total flux magnitude in the pre- and post-trifurcation regions of the separator. The small magnitude of the Brownian diffusion flux seen in Figure 13(c) confirms that its contribution to the total flux within the device is negligible. Figure 13(d) depicts the magnitude of the sum of the migration fluxes. The total migration flux is large in the separator zone when compared to the sections before and the trifurcation region, and is also large when compared to the upstream of the separator arms. The total flux at any section is the sum of the oppositely oriented individual fluxes in terms of the movement towards the centre, significantly lowering the net flux magnitude. This trend is important in particle migration from high-shear to low-shear zones, since it enables the particles to reach an equilibrium distribution in concentration. Hence, particles accumulate in regions with lower shear stress, i.e., in the central core of a microchannel. In the trifurcation zone, the flow is distributed in the separate arms, altering the local strain rate, thus affecting the nature of individual fluxes, in turn, the total diffusive flux, and the hematocrit distribution.

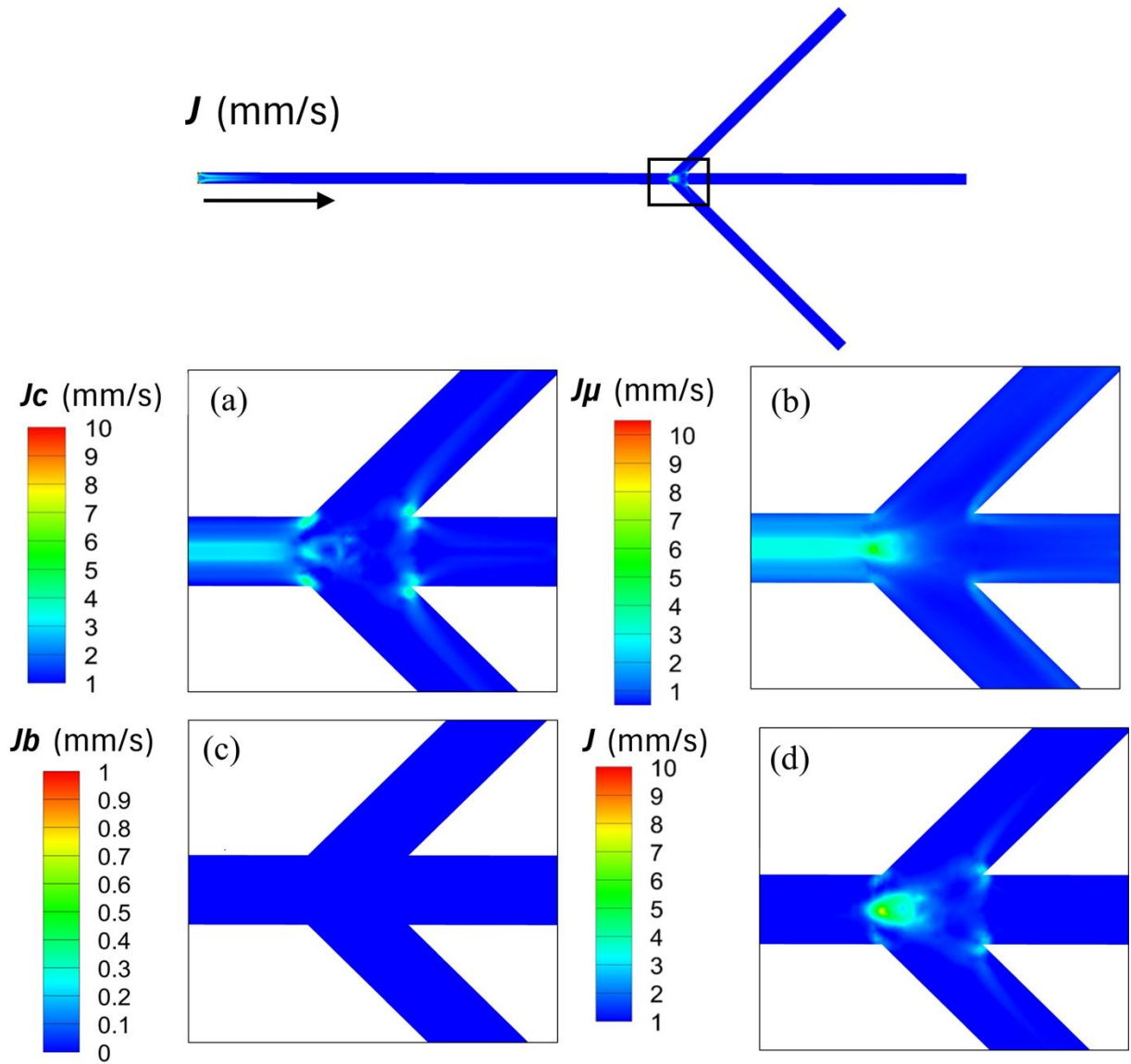

Figure 13: Contours depicting the magnitude of the migration fluxes at the midplane of the trifurcated microchannel with $H = 60$ µm, $\alpha = 45°$ for $Q = 0.25$ ml/min and average $\phi = 0.4$ on the inflow plane. Fluxes considered are: (a) collision flux $\vec{J}_c$, (b) viscosity gradient flux $\vec{J}_\mu$, (c) Brownian flux $\vec{J}_b$, and (d) total flux contours of $|\vec{J}| = |\vec{J}_c + \vec{J}_\mu + \vec{J}_b|$. The complete midplane of the microchannel is first shown for the total flux, and is followed by a magnified view of the indicated frame.

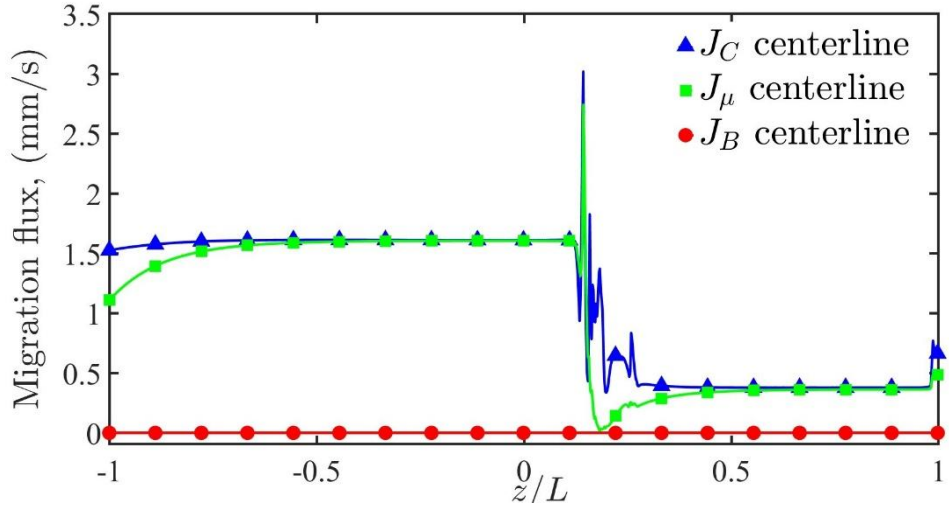

Figure 14. Variation of the centreline diffusive fluxes in the axial direction shown for $H = 60$ µm, $\alpha = 45°$, and $Q = 0.25$ ml/min. The magnitudes of the collision flux $\vec{J}_c$, viscosity gradient flux $\vec{J}_\mu$ and the Brownian flux $\vec{J}_b$ are shown on the centreline of the microchannel. Large changes in the fluxes occur in the trifurcation region of the separator ($z=0$).

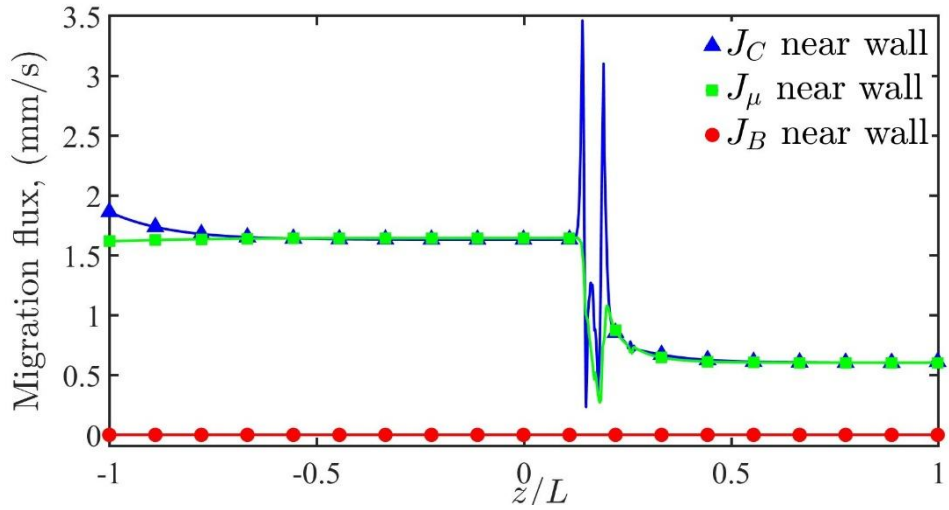

Figure 15: Variation of the near-wall diffusive fluxes in the axial direction shown for $H = 60$ µm, $\alpha = 45°$, and $Q = 0.25$ ml/min. The magnitudes of the collision flux $\vec{J}_c$, viscosity gradient flux $\vec{J}_\mu$, and the Brownian flux $\vec{J}_b$ are shown near the wall of the microchannel. Large changes in the fluxes occur in the trifurcation region of the separator ($z=0$).

Figures 14 and 15 are plots of the magnitudes of the three fluxes at the centreline and near the wall, respectively. It is observed that the collision and viscosity-related fluxes contribute significantly in terms of magnitude but are oppositely oriented in the straight segments of the channel. In comparison, Brownian diffusion has a negligible contribution. The diffusivity of RBC in plasma is a small quantity, leading to a Brownian flux of the order of $10^{-5}$ (m/s). In comparison, the collision and viscosity fluxes are of the order of $10^{-2}$ and $10^{-3}$ (m/s), thus being much larger. Figures 14 and 15 show that the collision flux is consistently larger than the one related to the viscosity gradient over the length of the separator. In the trifurcated region, the flow is distributed among the separator arms and central channel, and the component fluxes are not strictly opposed anymore. Hence, the total flux magnitude is greater when compared to the pre- and post-trifurcated regions, causing the hematocrit distribution to be non-uniform.

Major factors responsible for the diffusive flux behaviour in the geometry are summarized in Figure 16. These factors include the gradients of the hematocrit and strain rate magnitude, along with the gradient of the apparent viscosity (Equations 8-10). The contours of strain rate and the gradient of strain rate-hematocrit product are shown in Figure 16(a) and Figure 16(b). While the strain rate is large near the walls in the straight sections of the channels, the gradient quantities are redistributed in the trifurcation region. The behaviour of the gradient function affects the distribution of the collision fluxes, as seen in the similarity of Figure 16(b) and Figure 16(c). Apparent viscosity and its gradient contours are shown in Figure 16(d) and Figure 16(e), respectively, which create a viscosity flux (Equation 9) and ensure its redistribution, as in Figure 16(f). In the incoming channel, shear rates are small in the core, while apparent viscosity is large, and the gradient in viscosity is quite large. The third factor is responsible for a large viscosity flux in the incoming channel directed from the axis to the wall. Lowering the flow rate and higher hematocrit in the main outgoing channel lowers viscosity flux, and the particle migration is then determined by the collision flux alone. The collision-induced flux makes the particle migrate from a high shear rate region to a low shear rate region, whereas the varying viscosity flux ensures the particle migrates from a high concentration to a low concentration region. Hence, in the main incoming channel, the collision and viscosity fluxes are of opposite signs, creating a possibility of a stable particle distribution. In the trifurcation region, variations in the strain rate of the base flow affect the relative magnitudes and directions of the mass fluxes. Both fluxes are strongly influenced by the local shear rate and hematocrit distribution, which are sensitive to the flow and geometrical parameters. Consequently, alterations in the device geometry directly modify the local shear rate, the viscosity distribution, and the principal diffusive fluxes and ultimately, the overall separation effectiveness.

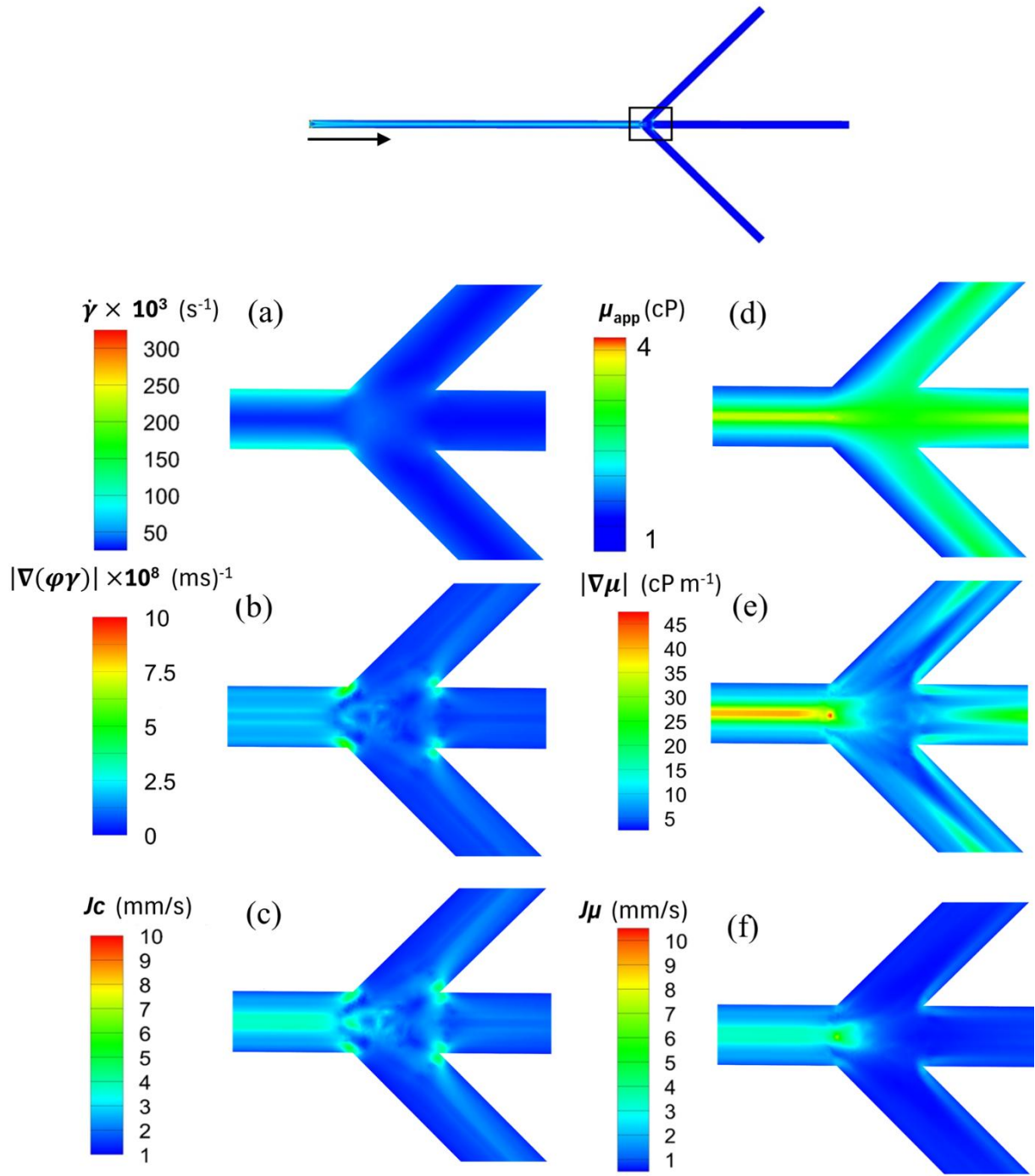

Figure 16: The magnitude contours (a) strain rate, (b) gradient of strain-rate hematocrit product, and (c) collision flux $\vec{J}_c$ are shown. Similarly, (d) apparent viscosity, (e) viscosity gradient, and (f) viscosity flux $\vec{J}_\mu$ are presented.

## 3.3. Effect of hematocrit dilution

According to the Fahraeus-Lindqvist effect [4], particles migrate away from the wall, toward the central axis of the channel. The migration towards the central axis develops a zone of depleted hematocrit near the wall. This region is referred as the cell-free layer. Developing a cell-free layer (CFL) is important in the RBCs-platelets separation process within the plasma. In a trifurcation separator studied here, larger cells such as RBCs follow the Fahraeus effect, and an RBC-free region is formed near the wall. In a simultaneous process called margination, platelets, being significantly smaller, migrate toward the wall in the opposite direction. The development of a platelet-rich cell-free layer and the extraction of plasma from this zone is the central purpose of the separator device. At the same time, the hematocrit concentration continuously increases in the main channel. In summary, blood collected in the separator arm has less hematocrit and is high in platelet concentration.

The mathematical model of the present work is focused on the RBC concentration since it is the dominant particulate component in the plasma. The enrichment of plasma by platelets in the separator arm is measured in terms of an increase in the RBC content in the main channel, just downstream of the trifurcation zone. Alternatively, the increase in CFL layer thickness is also a measure of the separator's effectiveness.

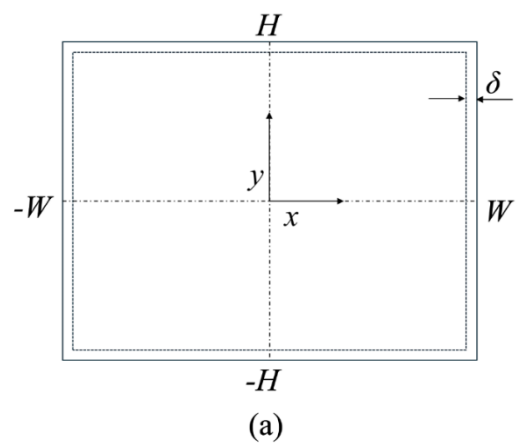

(a)

Figure 17: Schematic of the rectangular cross-section of the separator device followed in the CFL calculation with relevant geometric dimensions shown.

To assess the separator performance, two quantities of interest are the tube hematocrit $\phi^T$ and the discharge hematocrit $\phi^D$. These are defined by the distributions of the local hematocrit $\phi(x,y)$ and velocity $u(x,y)$ for a channel with a rectangular cross-section, as follows (Figure 17):

$$4WH\phi^T = \int_{-W}^{W}\int_{-H}^{H}\phi(x,y)dydx \tag{18}$$

$$4WH\overline{U}\phi^D = \int_{-W}^{W}\int_{-H}^{H}u(x,y)\phi(x,y)dydx \tag{19}$$

Here, $\overline{U}$ is the average velocity for the channel of a rectangular cross-section, given as

$$\overline{U} = \frac{\int_{-W}^{W}\int_{-H}^{H}u(x,y)dxdy}{4WH}$$

Since velocities are larger at the core of the channel, the difference between the two average hematocrit values $(\phi^T - \phi^D)$ may be recognized as an indicator of the cell-free layer (CFL) thickness at the vessel wall. Studies in a straight tube show that the CFL thickness becomes more pronounced as the tube diameter decreases. This result is the origin of the Fahraeus–Lindqvist effect, which describes the reduction in apparent blood viscosity over a cross-section with decreasing tube diameter, as observed in blood flow experiments using glass tubes [20]. It is directly related to the thickness of the CFL layer.

The CFL layer ($\delta$) can be calculated for a channel of rectangular cross-section as follows (Figure 17):

$$\delta(H+W)\overline{U}\phi^D = \int_{-W}^{W}\int_{-H}^{H}u(x,y)(\phi(x,y)-\phi^T)dydx \tag{20}$$

Using Equations 18 and 19 in Equation 20, the thickness of the cell-free layer is obtained as:

$$\delta = L_c\left(1 - \frac{\phi^T}{\phi^D}\right) \tag{21}$$

Here $L_c$ is the characteristic length, given as $\frac{4WH}{W+H}$ for the rectangular cross-section.

The definition of the cell-free layer is applied to the trifurcated microchannel of square cross-section and compared with the trends for a straight channel, Figure 18. In Figure 18(a), the microchannels are of width $H = 60$ μm, the cross-section being a square. The separator arms are at an angle of $\alpha = 45°$. The motivation behind parametrically studying the effect of inlet hematocrit concentration is to examine the effect of dilution of blood on the separator performance. From Figure 18(b), it is observed that as the inlet RBC concentration increases, the CFL thickness decreases, depicting a higher concentration of RBC near the wall. In contrast, at low inlet concentrations, a higher CFL layer thickness is observed. This result suggests that dilution is helpful in lowering wall hematocrit concentration, leading to platelet-enriched plasma near the walls. However, the changes in flow rate do not impact the CFL thickness, as seen in Figure 18(c), consistent with the velocity distribution presented in earlier sections.

Figure 18(d) presents a similar analysis of dilution in a trifurcated microchannel with $H = 60$ and 80 µm in terms of effectiveness. Here, separation effectiveness is calculated as the percentage difference between the discharge hematocrit at the inlet and the major outlet (2) of the microchannel. The smaller channel opening consistently delivers higher effectiveness over the range of inlet hematocrit levels studied, though both channels show better performance at higher levels of dilution. The superiority of the smaller channel is aligned with the outcome of Section 3.1, where a greater degree of separation was possible in smaller microchannels.

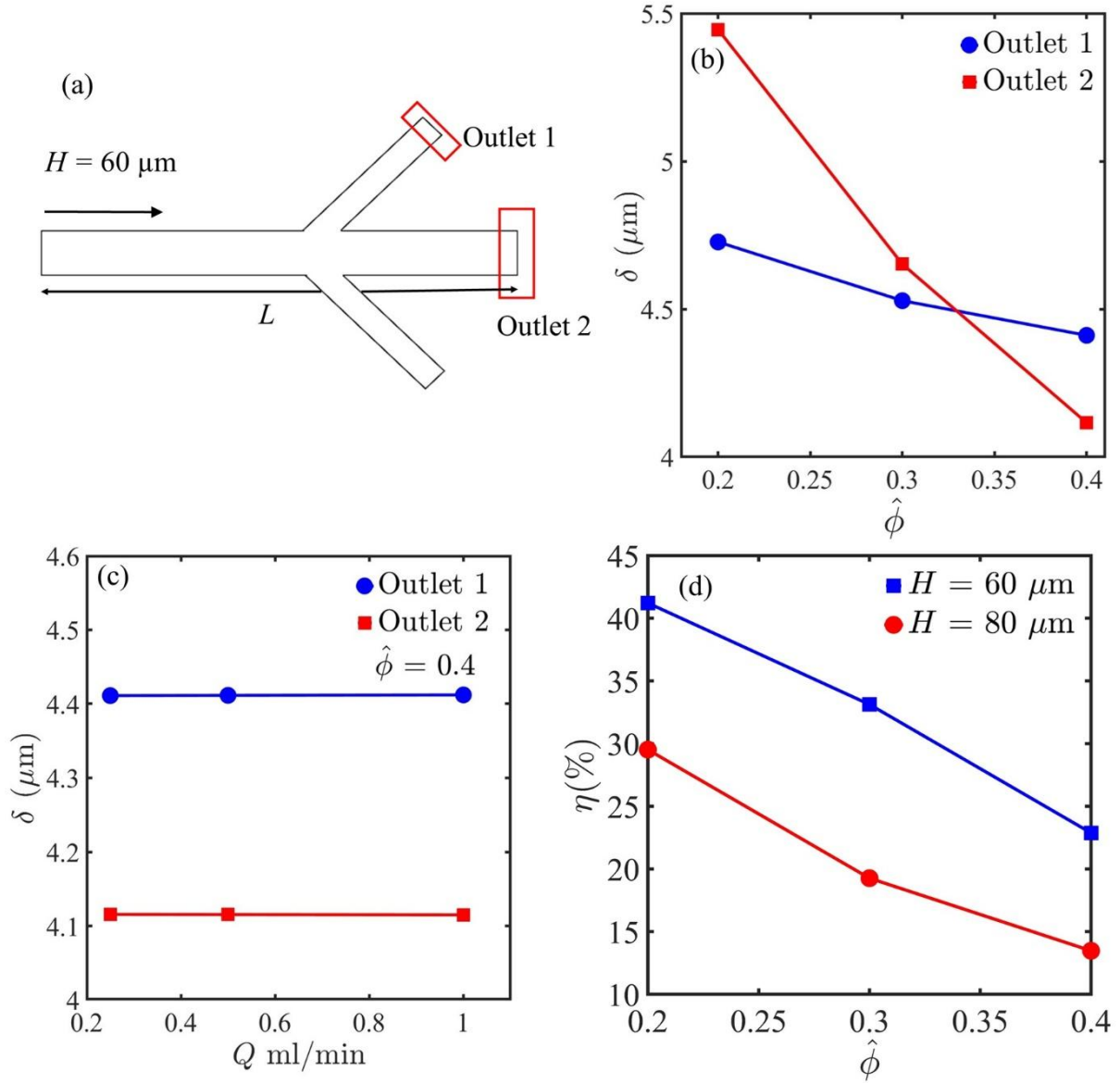

Figure 18: (a) Trifurcated microchannel considered for the calculation of the cell-free layer thickness. (b) Variation of CFL thickness as a function of the average hematocrit in the microchannel with $H = 60$ µm and inlet flow rate $Q = 0.25$ ml/min; (c) CFL thickness at selected flow rates for inlet hematocrit $\hat{\phi} = 0.4$; and (d) Effectiveness of the trifurcation device from 2D simulations for $H = 60$ and $80$ µm with $45°$ separator arms.

### 3.4. Effects of bulk temperature on hematocrit separation

Control of temperature during flow in a trifurcated microchannel for generating platelet-enriched plasma by separating RBCs is important. Ideally, the process should occur at the body temperature of 37°C since the stability of the constituents of blood is most assured under these conditions. Results presented in earlier sections assumed this temperature (Table 2). The sensitivity analysis of the device performance with temperature is shown next.

The rheology model for blood proposed by Apostolidis and Beris[18] adopts an Arrhenius-type functionality with temperature. The resulting viscosity variation with the strain rate magnitude is shown in Figure 19 for temperatures of 298 K and 310 K. A 10-30% difference in viscosity is seen here, with viscosity being lower at the higher temperature. Keeping the particle migration parameters such as collision frequency unchanged (Equations 8 and 9), the separator effectiveness at these two temperatures is explored further. The diffusion flux arising from the shear gradient remains unaffected, while terms involving viscosity are impacted. For the comparative study at two temperatures, the central channel width is taken to be 60 µm, and the separator arms are oriented at 45° from the main flow direction.

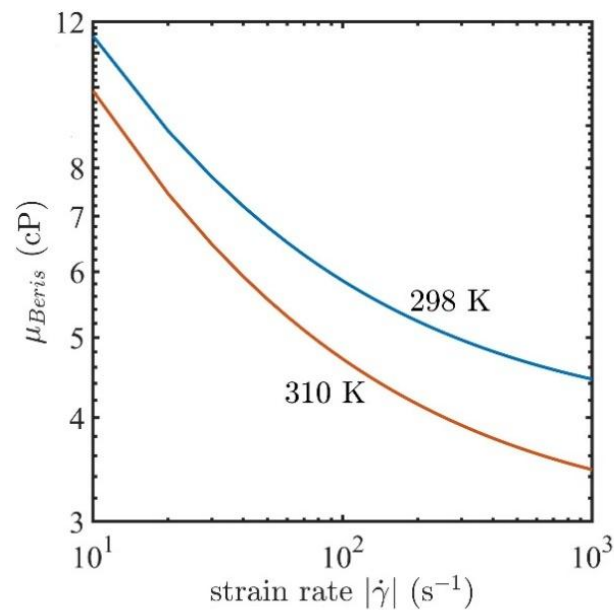

Figure 19: Blood viscosity as a function of the strain rate magnitude at temperatures of 298 K and 310 K [22].

The viscosity variation, along with the hematocrit and flux distribution, is shown in Figure 20. Figure 20(a) shows higher viscosity at the lower temperature (298 K), when compared to the distribution at body temperature of 310 K. Figure 20(b) shows the hematocrit distribution contours at the two temperatures, and no significant differences are observed.

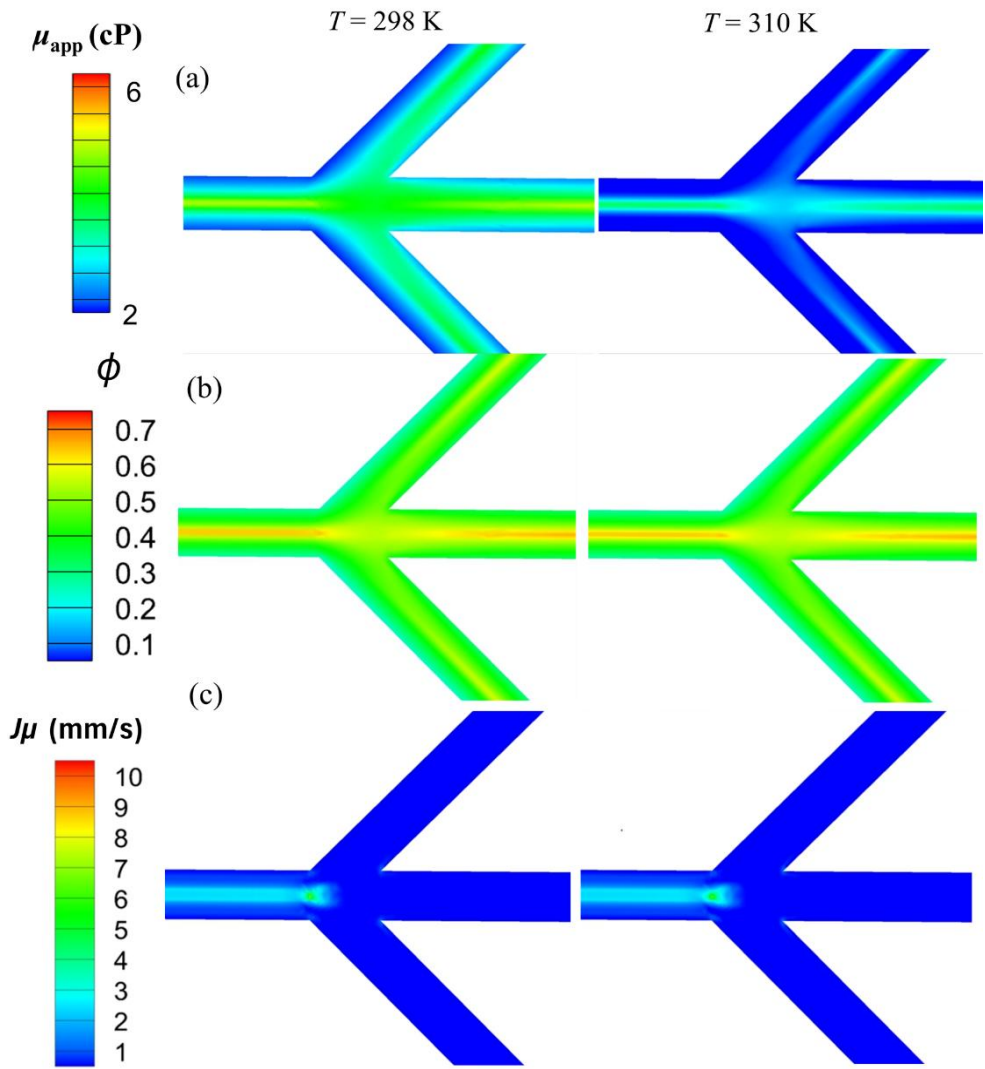

Figure 20: For blood with temperatures $T = 298$ K and 310 K, contours of (a) viscosity, (b) hematocrit $\phi$, and viscosity gradient flux magnitude $\vec{J}_\mu$ are shown.

Figure 20(c) shows the distribution of mass flux arising from a viscosity gradient. As in Figure 20(b), differences are negligible, confirming only a minor effect of temperature on the concentration distribution. The hematocrit distribution depends on the migration fluxes, which are a function of strain-rate magnitude, viscosity distribution, and hematocrit gradient function. If the temperature is spatially constant, no new gradient in viscosity is created. Viscosity gradient flux depends primarily on the strain rate magnitude, while contributing to the migration fluxes, and is unaffected by a uniform temperature field. Hence, the choice of the bulk temperature of the material does not affect the separator effectiveness. On the other hand, the stability of the blood constituents is temperature-dependent, and the choice of temperature should be made on medical considerations.

## 3.5. Effects of Constriction and Inlet Extension

Separation devices with changes in the channel design near the trifurcation point have been reported in the literature. An experimental study by Maria et al.[24] introduced a constriction in the microchannel to increase plasma recovery. This suggestion is examined in the present section. We introduce a constriction before the separator arm and compare the hematocrit discharge with the original microchannel. The dimensions of the constriction and its design are adopted from the reported study[24]. A point of difference is that the experimental setup is a collection of 10 micro-devices in series, while the analysis presented here is for a single unit. A schematic of the microchannel with a constriction is shown in Figure 21, where the inlet length $L_1$ = 1000 microns and a uniform depth of 50 microns. The average inlet hematocrit in simulations is 0.4, and the flow rate is 2 ml/h.

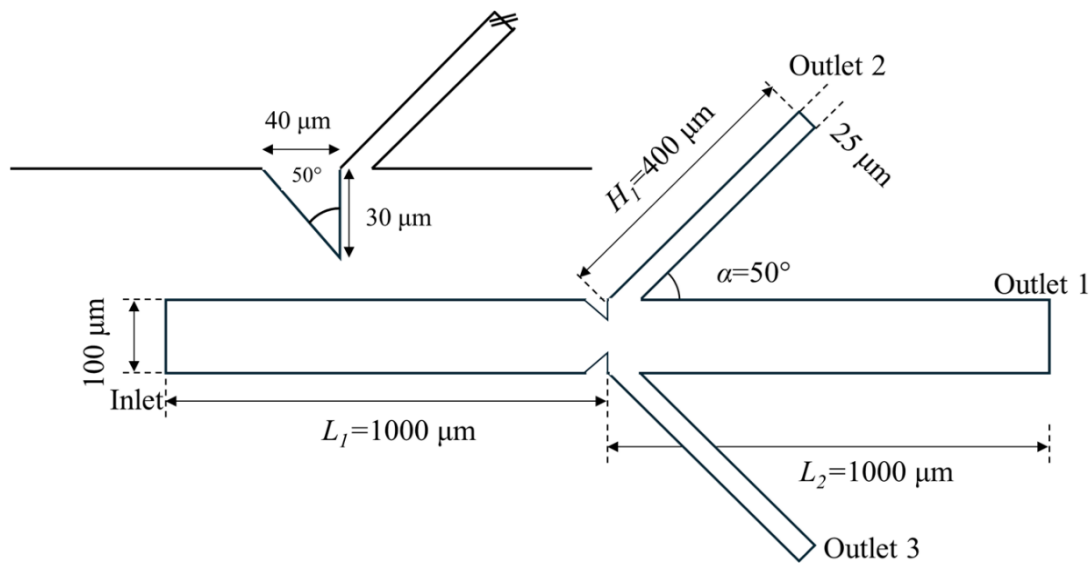

Figure 21: Separator device schematic with constriction whose dimensions are highlighted.

Since the experimental results use real blood samples and the geometry is significantly different, a direct comparison with simulations is not possible. Results are shown in terms of the difference in the hematocrit discharge in the main channel and the separator arms with respect to the inlet. The discharge hematocrit is calculated as per the discussion in Section 3.3 and is shown in Figure 22. The presence of a constriction is marked 'I', and a plain channel is denoted as 'II'. The two configurations are compared for two different entrance lengths. In Figure 22, the hematocrit discharge is shown at three locations of the microchannel without an inlet extension (Figure 22(a)) and with an inlet extension (Figure 22(b)). It is observed that with the introduction of the constriction, the value of hematocrit discharge in the separator arm is higher than in the separator arm of the microchannel without constriction, indicating that more hematocrit goes into the separator arm. However, the discharge is similar in the downstream section of the main channel. This observation indicates that the introduction of the constriction is unable to increase the hematocrit discharge in the main channel or decrease the hematocrit in the separator arm of the microchannel. To improve the channel effectiveness further, the inlet channel (before the separator arm) is extended to twice the original length, i.e., $2L_1$. The hematocrit discharge in the separator arm is now seen to be smaller in comparison to the non-extended version of the microchannel. The extension allows the hematocrit to migrate further toward the axis before arriving at the separator arm. The RBC concentration near the wall is lowered, which decreases the hematocrit discharge in the separator outlet.

The percentage difference in hematocrit discharge for the main and separator arm outlet with respect to the fully developed inlet concentration is shown in Figure 23. The shorter inlet section is considered in Figure 23(a), while the effect of the extension is shown in Figure 23(b) for a flow rate $Q = 2$ ml/hr and an inlet hematocrit of $\phi = 0.4$. In Figures 23(c) and (d), a balance in flow and hematocrit discharge confirms mass conservation. The flow and hematocrit discharges are respectively defined as

$$\begin{aligned} Q^D &= \int \vec{u} \cdot \vec{dA} \\ \phi^D &= \int \phi \vec{u} \cdot \vec{dA} \end{aligned} \quad (22)$$

Positive changes show an increase, and negative percentages depict a reduction in the concentration. Figure 23 confirms a positive impact of the inclusion of the inlet extension, whereas the constriction has a negligible effect in diminishing concentration in the separator arm.

The difference in trends between experiments[24] and simulation on the effect of a constriction on separation efficiency and plasma enrichment can be explained as follows. Within the framework of a diffusive flux model, particles migrate toward the axis; however, the cell-free layer does not have a zero hematocrit concentration. The reason is the continuum framework adopted for modeling particle migration. In a real microfluidic device, the plasma is depleted of RBC particulates in the wall region, and its concentration is quite low. Thus, the real separator performance is likely to exceed what is predicted by the diffusive flux framework and also reveal a beneficial effect of providing a constriction.

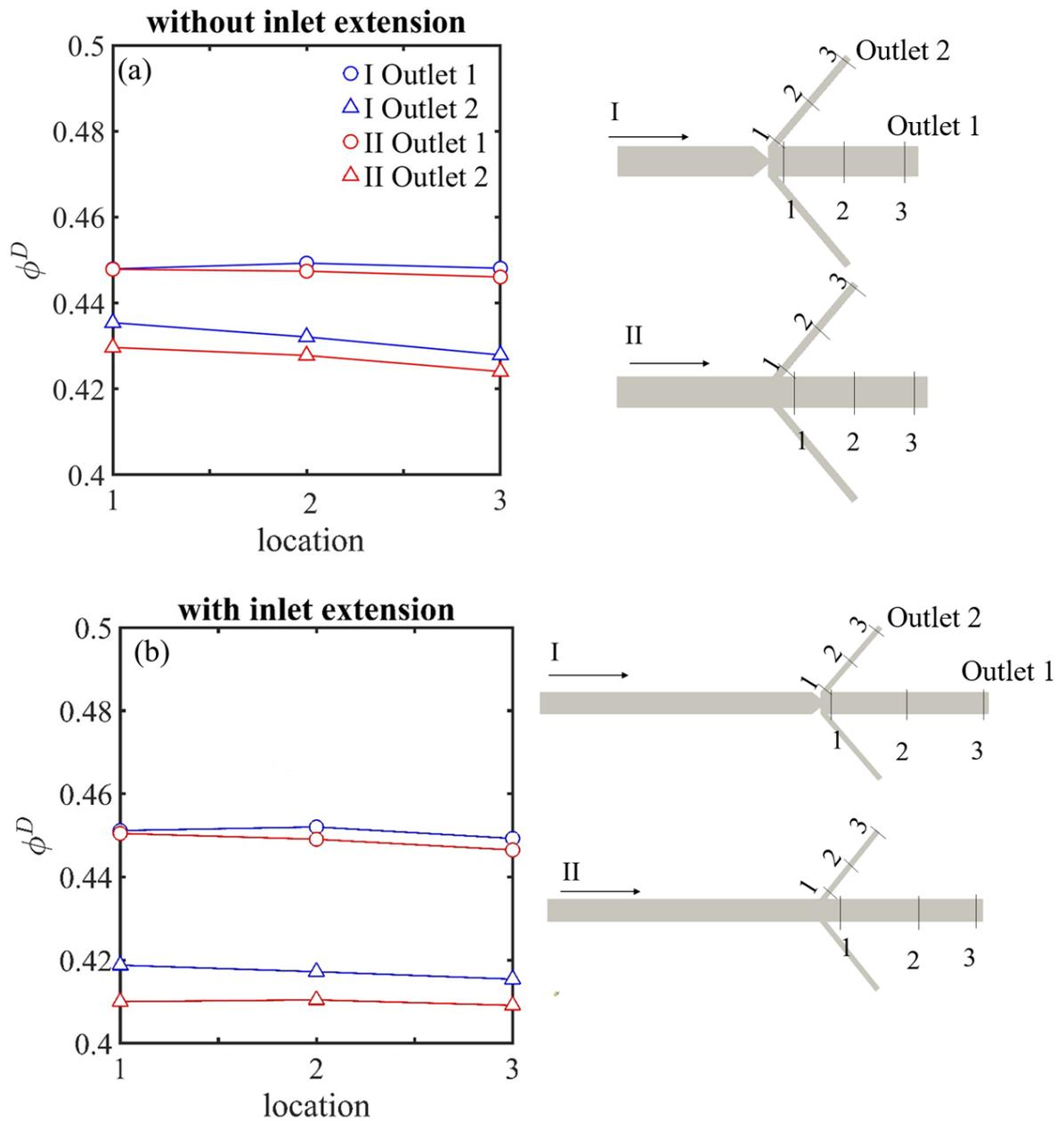

Figure 22: Hematocrit discharge calculated for both the microchannel with constriction (I) and without constriction (II), for (a) without inlet extension and (b) with inlet extension. Outlet 1 is that of the main channel, while Outlet 2 is that of the separator. Locations 1-3 are marked on the *x*-axis.

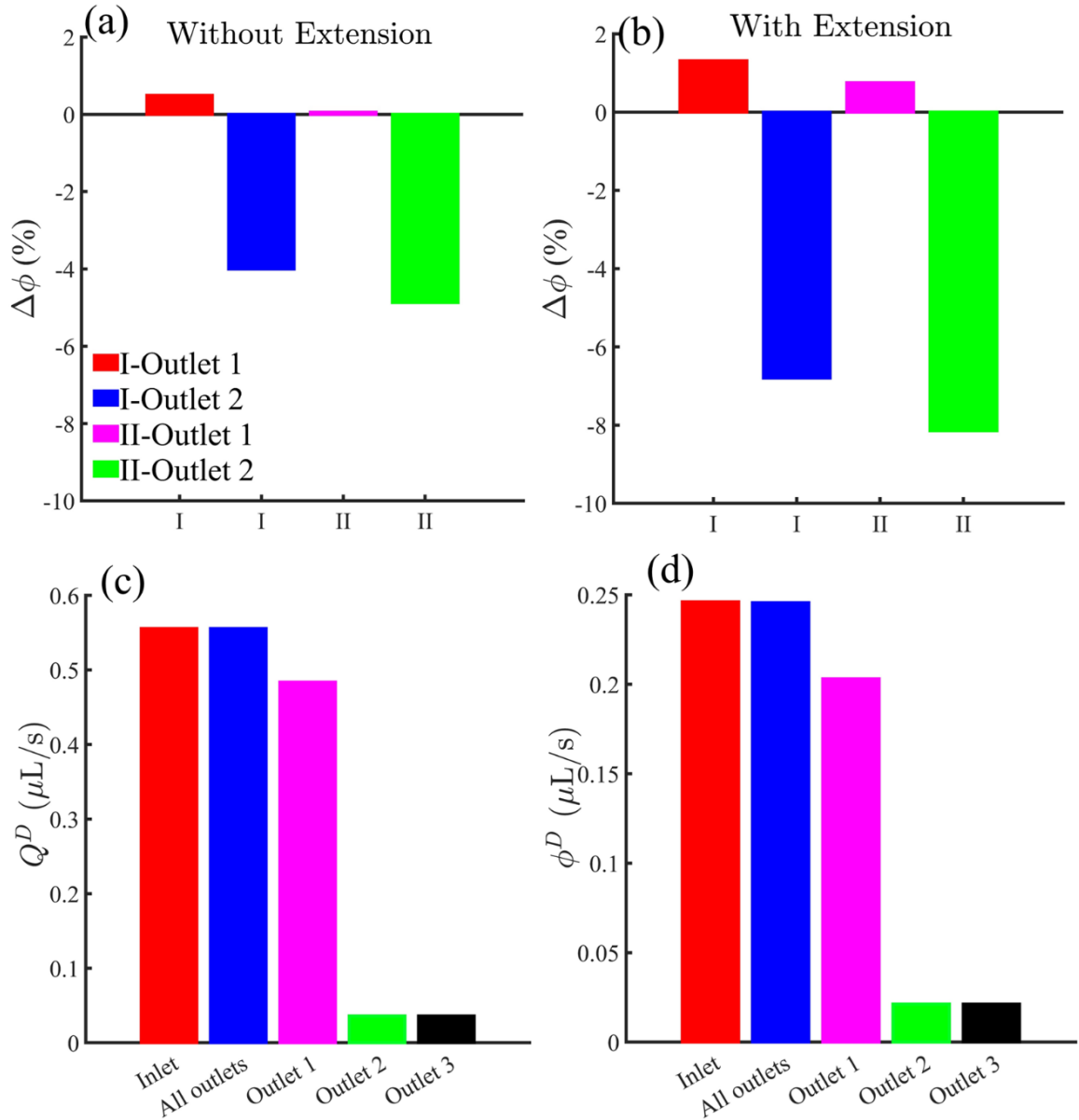

Figure 23 The percentage difference in hematocrit discharge for the separator device (a) without extension, and (b) with extension, including constriction shown for a flow rate of 2 ml/h and inlet hematocrit $\phi = 0.4$. Balances in (c) volumetric flow rate and (d) mass of species are shown for an inlet extended constricted (I) device. In (c) and (d), the inlet value matches the sum of all outlets within 0.2 %.

### 3.6. Coupling and decoupling of diffusive flux model (DFM) in 3D and 2D separator channels

While 3D simulations reveal completely the flow and mass transport phenomena, microfluidic simulations require fine discretization and correspondingly small-time steps, making the computational cost quite high. In the present work, a flow simulation of 1 second required a time step of 1 µs on the converged grid. Thus, it is reasonable to ask if 2D simulations can be used as a substitute, at least for testing the solver and performing parametric studies. Such studies are also useful in the study of wide microchannels, where ($W>>H$), where the out-of-plane velocity component is minimal. In the present section, 2D and 3D simulations are compared in a trifurcated channel and shown in Figures 24(a) and (b). The channel height is $H = 80$ µm, $\alpha = 45°$, and the inflow is 0.25 ml/min. In Figures 24(a) and (b), the velocity profile is normalised by the average velocity, and the hematocrit profile is shown with normalised $y$-coordinates before and after the junction. It is observed that velocity profiles are comparable, but the concentration profile shows significant differences. This trend is explainable because of the absence of wall effects in the 2D approach. Another comparison is performed to study the effects of channel width. Here, the 2D trifurcation microchannel and a 3D square channel of the same characteristic length are compared for a flow rate of 0.25 ml/min. Figure 24(c) shows the cell-free layer with respect to the channel width ($H$) and compares the 2D trifurcation microchannel with the 3D straight square channel for the selected main channel width. It is seen that the channel width is a crucial geometrical parameter for understanding separation effectiveness, as confirmed by the decrease in the CFL layer in both simulations.

A second approach to improve the speed of calculations is to selectively decouple the momentum and the species transport equations. Here, the flow distribution is first calculated by solving the continuity and momentum transport equations using the blood viscosity model with an average hematocrit concentration of 0.4. Using the velocity distribution, the species transport equation is solved separately, thus decoupling the two equations. Figure 25 compares the predictions of the coupled and decoupled solvers for the trifurcated separator device of $H = 60$ µm, $H' = 60$ µm, and angle $\alpha = 45°$ with the inflow rate 0.25 ml/min. Both two and three-dimensional simulations are compared.

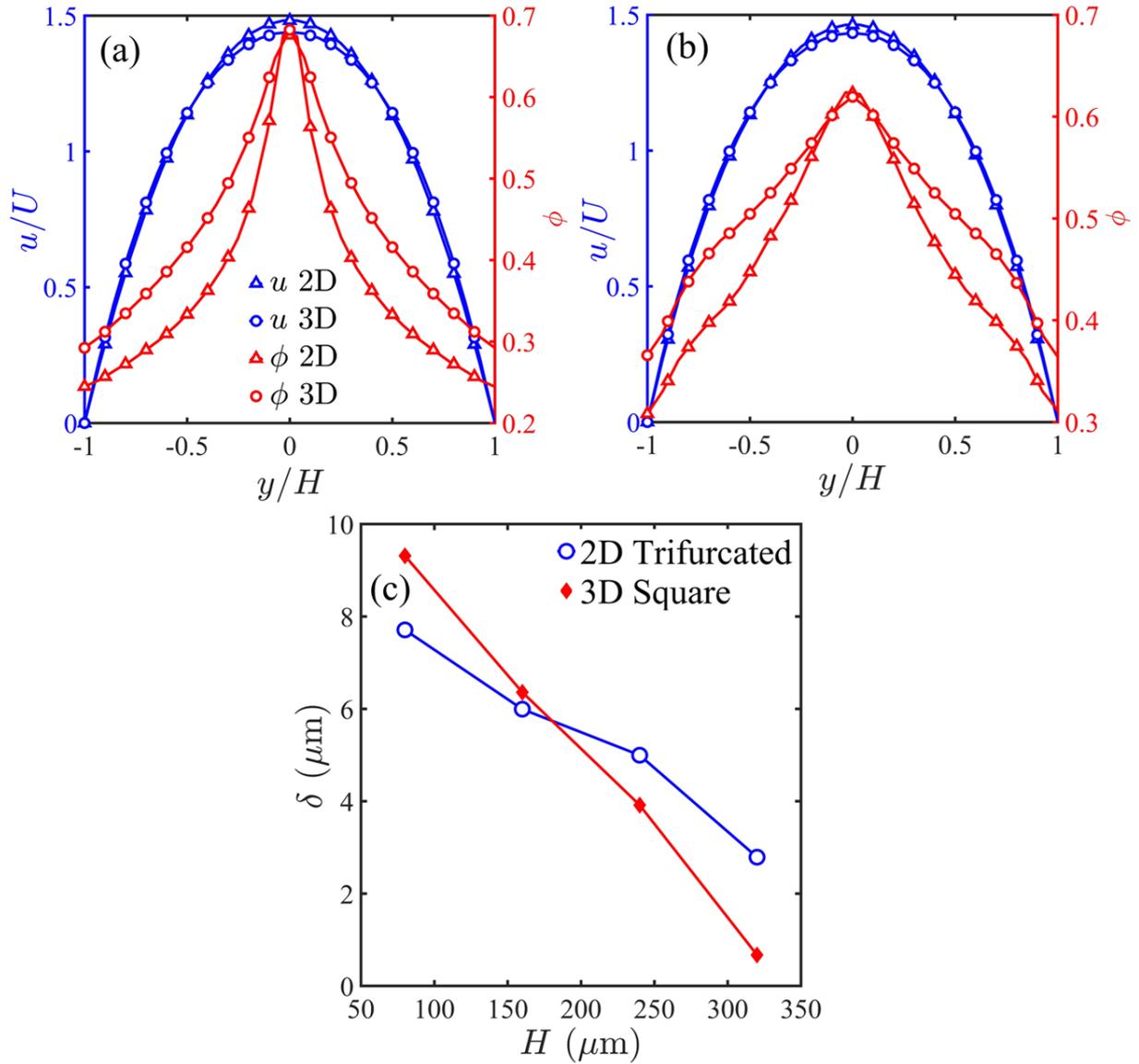

Figure 24: Normalised velocity magnitude ($U$ is average velocity) and hematocrit concentration are shown with the normalised $y$ coordinate for (a) prior to trifurcation and (b) post-trifurcation of $H=80$ μm and $\alpha = 45°$ for 2D and 3D analysis of the separator device. Part (c) compares the CFL thickness for 2D trifurcated microchannel with a 3D square channel for various main channel widths ($H$).

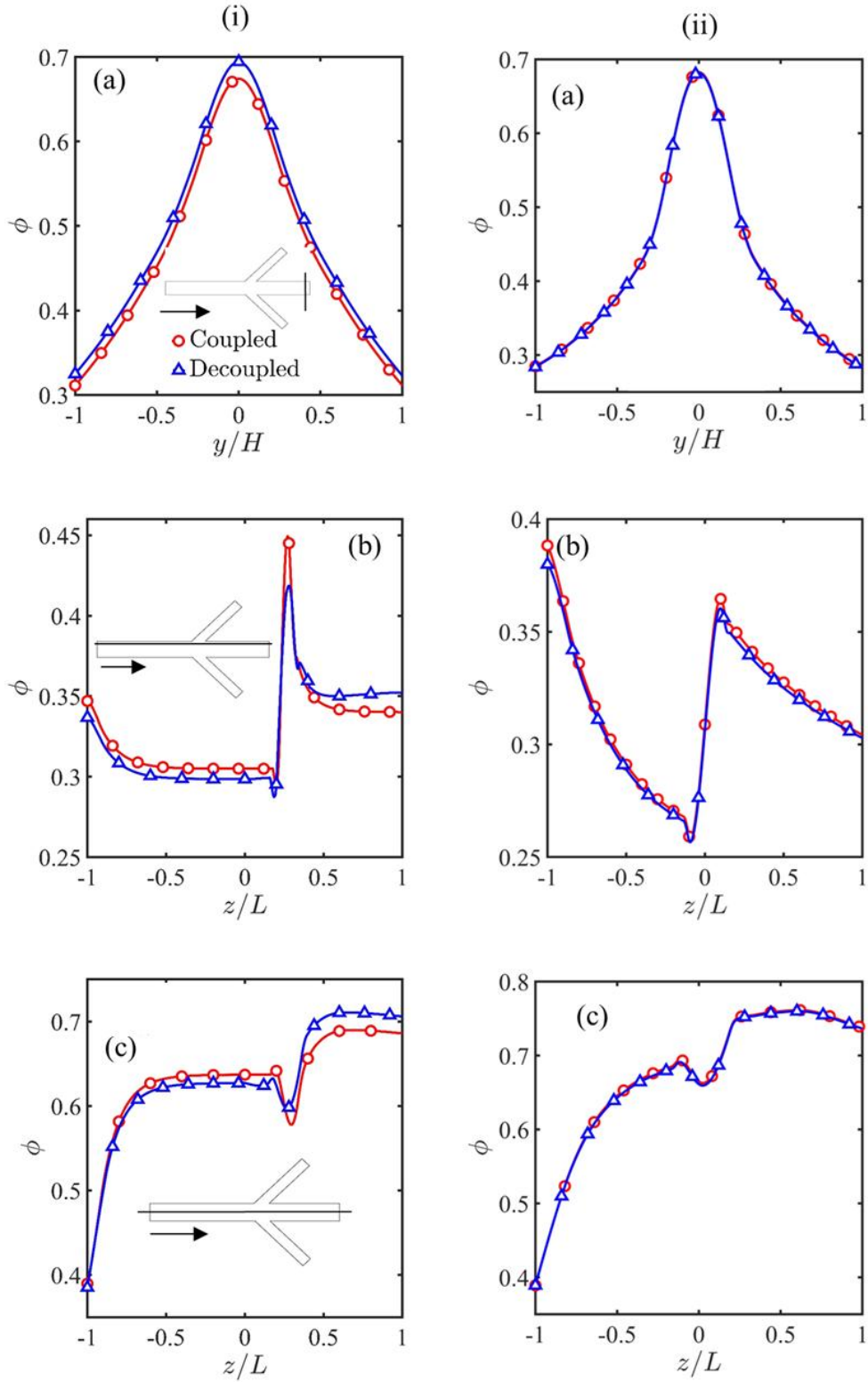

Figure 25 Hematocrit concentration at the (a) outlet, (b) near wall, and (c) at the centre of the microchannel for (i) 3D microchannel of $H = 60$ μm with (ii) 2D 80 μm microchannel when the separator angle is 45°. Coupled and decoupled versions of the OpenFOAM solver are compared.

In Figures 25 (i) and (ii) (left and right columns), the concentration profiles are shown for 2D and 3D simulations, respectively, for momentum-species transport equations coupled and decoupled. In Figure 25, the location considered for plotting the concentration profile is shown in the inset of the figure. The profiles shown in the figure are the outlet, Figure 25(a), near the wall, Figure 25(b), and the centreline of the microchannel, Figure 25(c). The coupled and decoupled concentration profiles in 3D simulations appear to be close to each other, while for 2D, they are practically identical. These trends broadly confirm that the decoupled approach can be utilized to derive the hematocrit concentration distribution.

## 4. Conclusions

Simulations and analysis of a trifurcated microchannel device for passive separation of platelet-enriched plasma and RBCs in blood are carried out in the present study. The diffusive flux model is included in the modelling of flow and transport to account for hematocrit migration across the channel. A rheology model specifically developed for blood has been adopted to determine the local values of apparent viscosity. The performance of the trifurcated microchannel, namely its effectiveness, is studied with respect to geometrical parameters, flow rates, and bulk temperature. The relative contribution of the components of the composite diffusive flux is examined. The full 3D solutions are also compared with a simplified 2D treatment in terms of the effectiveness of the separator device, and the results were seen to be qualitatively similar. Specific conclusions arrived at in the present work are summarized below.

1. In terms of geometry, the cross-sectional area of the separator channel opening is seen to be more significant than the branching angles in determining the separator effectiveness. Smaller apertures are seen to deliver higher effectiveness and are to be preferred. In this context, note that the separator arm width used in experiments[23] is 3 times lower than the main width of the main channel.
2. The inlet flow rate affects the velocity distribution, but for the geometry studied, the species concentration distribution remains practically unchanged. Accordingly, cell-free layer thickness and separation effectiveness are unchanged for the range of flow rates studied.
3. With increasing hematocrit concentration, the CFL thickness and separation effectiveness decrease, suggesting that a dilute sample at the inlet of the separator device is expected to yield higher performance.

4. Despite the change in viscosity, the hematocrit separation in the device was not altered by changes in bulk temperature between 25°C (room temperature) and 37°C (body temperature).
5. Among the fluxes studied, the collision flux and the one related to the gradient in apparent viscosity flux are seen to be important in determining the rate of migration of the RBCs. In comparison, the flux associated with Brownian motion is significantly smaller in terms of the order of magnitude.
6. The inclusion of the inlet length extension in the separator arm helps in improving separator effectiveness, whereas a constriction placed at the trifurcation did not show any improvement. The second result contradicts experiments reported in the literature and will require better modeling of the wall boundary condition for the species transport equation.
7. A decoupled approach employing a two-dimensional numerical framework has been shown to be effective for predicting hematocrit distribution and assessing separator device performance with respect to geometrical and flow parameters.

**Acknowledgment**

Authors acknowledge the National Supercomputing Mission (NSM) for providing computing resources of PARAM SANGANAK at IIT Kanpur (India) supported by the Ministry of Electronics and Information Technology (MeitY) and Department of Science and Technology (DST), Government of India. Also, we would like to thank the computer centre (www.iitk.ac.in/cc) at IIT Kanpur for providing the resources to carry out the reported work. The authors gratefully acknowledge the kind support provided by the DST NSM project with a sanction number DST/NSM/R&D HPC-Applications/2021/18.

**Author Declarations**

The authors have no conflicts to disclose.

**References**

[1] A.S. Popel, and P.C. Johnson, "MICROCIRCULATION AND HEMORHEOLOGY," Annu. Rev. Fluid Mech. **37**(1), 43–69 (2005).
[2] R. Yen, and Y. Fung, "Effect of velocity distribution on red cell distribution in capillary blood vessels," American Journal of Physiology-Heart and Circulatory Physiology **235**(2), H251–H257 (1978).
[3] Y.-C. Fung, "Stochastic flow in capillary blood vessels," Microvascular Research **5**(1), 34–48 (1973).

[4] K. Svanes, and B.W. Zweifach, "Variations in small blood vessel hematocrits produced in hypothermic rats by micro-occlusion," Microvascular Research **1**(2), 210–220 (1968).
[5] R. Fåhraeus, "THE SUSPENSION STABILITY OF THE BLOOD," Physiological Reviews **9**(2), 241–274 (1929).
[6] R. Fåhræus, and T. Lindqvist, "THE VISCOSITY OF THE BLOOD IN NARROW CAPILLARY TUBES," American Journal of Physiology-Legacy Content **96**(3), 562–568 (1931).
[7] A.M. Robertson, A. Sequeira, and M.V. Kameneva, "Hemorheology," in *Hemodynamical Flows: Modeling, Analysis and Simulation*, (Birkhäuser Basel, Basel, 2008), pp. 63–120.
[8] R.T. Carr, and L.L. Wickham, "Plasma skimming in serial microvascular bifurcations," Microvascular Research **40**(2), 179–190 (1990).
[9] V. Laxmi, S. Tripathi, S.S. Joshi, and A. Agrawal, "Separation and Enrichment of Platelets from Whole Blood Using a PDMS-Based Passive Microdevice," Industrial & Engineering Chemistry Research **59**(10), 4792–4801 (2020).
[10] S. Tripathi, A. Prabhakar, N. Kumar, S.G. Singh, and A. Agrawal, "Blood plasma separation in elevated dimension T-shaped microchannel," Biomedical Microdevices **15**, 415–425 (2013).
[11] C. Truesdell, and C. Truesdell, *Historical Introit the Origins of Rational Thermodynamics* (Springer, 1984).
[12] M. Massoudi, "A note on the meaning of mixture viscosity using the classical continuum theories of mixtures," International Journal of Engineering Science **46**(7), 677–689 (2008).
[13] M. Massoudi, "A Mixture Theory formulation for hydraulic or pneumatic transport of solid particles," International Journal of Engineering Science **48**(11), 1440–1461 (2010).
[14] J. Kim, J.F. Antaki, and M. Massoudi, "Computational study of blood flow in microchannels," Journal of Computational and Applied Mathematics **292**, 174–187 (2016).
[15] K. Chandran, I.S. Dalal, K. Tatsumi, and K. Muralidhar, "Numerical simulation of blood flow modeled as a fluid-particulate mixture," Journal of Non-Newtonian Fluid Mechanics **285**, 104383 (2020).
[16] D. Leighton, and A. Acrivos, "The shear-induced migration of particles in concentrated suspensions," J. Fluid Mech. **181**(1), 415 (1987).
[17] R.J. Phillips, R.C. Armstrong, R.A. Brown, A.L. Graham, and J.R. Abbott, "A constitutive equation for concentrated suspensions that accounts for shear-induced particle migration," Physics of Fluids A: Fluid Dynamics **4**(1), 30–40 (1992).
[18] A.J. Apostolidis, and A.N. Beris, "Modeling of the blood rheology in steady-state shear flows," Journal of Rheology **58**(3), 607–633 (2014).
[19] P. Giri, K. Chandran, K. Muralidhar, and I. Saha Dalal, "Effects of coupling of mass transport and blood viscosity models for microchannel flows," Journal of Non-Newtonian Fluid Mechanics **302**, 104754 (2022).
[20] H. Lei, D.A. Fedosov, B. Caswell, and G.E. Karniadakis, "Blood flow in small tubes: quantifying the transition to the non-continuum regime," Journal of Fluid Mechanics **722**, 214–239 (2013).
[21] A. Yazdani, and G.E. Karniadakis, "Sub-cellular modeling of platelet transport in blood flow through microchannels with constriction," Soft Matter **12**(19), 4339–4351 (2016).
[22] X. Li, A.S. Popel, and G.E. Karniadakis, "Blood–plasma separation in Y-shaped bifurcating microfluidic channels: a dissipative particle dynamics simulation study," Physical Biology **9**(2), 026010 (2012).
[23] V. Laxmi, S. Tripathi, S.S. Joshi, and A. Agrawal, "Separation and Enrichment of Platelets from Whole Blood Using a PDMS-Based Passive Microdevice," Ind. Eng. Chem. Res. **59**(10), 4792–4801 (2020).

[24] M.S. Maria, B.S. Kumar, T.S. Chandra, and A.K. Sen, "Development of a microfluidic device for cell concentration and blood cell-plasma separation," Biomed Microdevices **17**(6), 115 (2015).

[25] A. Shauly, A. Wachs, and A. Nir, "Shear-induced particle migration in a polydisperse concentrated suspension," Journal of Rheology **42**(6), 1329–1348 (1998).

[26] G.P. Krishnan, S. Beimfohr, and D.T. Leighton, "Shear-induced radial segregation in bidisperse suspensions," J. Fluid Mech. **321**, 371–393 (1996).

[27] R.M. Miller, and J.F. Morris, "Normal stress-driven migration and axial development in pressure-driven flow of concentrated suspensions," Journal of Non-Newtonian Fluid Mechanics **135**(2–3), 149–165 (2006).

[28] I.M. Krieger, "Rheology of monodisperse latices," Advances in Colloid and Interface Science **3**(2), 111–136 (1972).

[29] M.K. Lyon, and L.G. Leal, "An experimental study of the motion of concentrated suspensions in two-dimensional channel flow. Part 1. Monodisperse systems," J. Fluid Mech. **363**, 25–56 (1998).